\documentclass[twocolumn]{aastex631}

\usepackage[utf8]{inputenc}
\DeclareUnicodeCharacter{2212}{-}
\usepackage{graphicx}
\graphicspath{{./}{./figures/}}
\usepackage{amsmath}
\usepackage{afterpage}
\usepackage{enumitem}
\usepackage{xcolor}
\usepackage{float}
\setenumerate{itemsep=0mm}
\usepackage{hyperref}

\definecolor{mypink}{cmyk}{0, 0.7808, 0.4429, 0.1412}
\definecolor{newblue}{cmyk}{1,0.7,0,0}
\definecolor{cadmiumgreen}{rgb}{0.0, 0.42, 0.24}
\definecolor{orchid}{rgb}{0.85, 0.44, 0.84}

\usepackage{amsmath}
\usepackage{amssymb}
\usepackage{xspace}
\usepackage{xifthen}
\usepackage{eso-pic}

\newcommand{\Gaia}{{\it Gaia}\xspace}

\definecolor{forestgreen}{HTML}{228B22}
\definecolor{urlblue}{HTML}{000000}

\newcommand{\eg}{e.g.\xspace}

\mathchardef\mhyphen="2D

\newcommand{\roughly}{\ensuremath{ {\sim}\,} }

\newlength{\dhatheight}

\newcommand{\code}[1]{\texttt{#1}\xspace}

\newcommand{\unit}[1]{\ensuremath{\mathrm{\,#1}}\xspace}

\newcommand{\degree}{\ensuremath{{}^{\circ}}\xspace}

\newcommand{\mas}{\unit{mas}}
\newcommand{\amin}{\unit{arcmin}}
\newcommand{\asec}{\unit{arcsec}}

\newcommand{\second}{\unit{s}}

\newcommand{\magn}{\unit{mag}}
\newcommand{\mmag}{\unit{mmag}}

\newcommand{\e}{\unit{e^{-}}}

\newcommand{\secref}[1]{Section~\ref{sec:#1}}
\newcommand{\appref}[1]{Appendix~\ref{app:#1}}
\newcommand{\tabref}[1]{Table~\ref{tab:#1}}

\newcommand{\figref}[1]{Figure~\ref{fig:#1}}

\newcommand{\var}[1]{%
 \ensuremath{\texttt{\MakeUppercase{#1}}}\xspace
}
\newcommand{\bandvar}[2][]{%
  \ifthenelse{\isempty{#1}}{\var{#2}}{\var{#2\_#1}}%
}
\newcommand{\spreadmodel}[1][]{\bandvar[#1]{spread\_model}}
\newcommand{\spreaderrmodel}[1][]{\bandvar[#1]{spreaderr\_model}}

\newcommand{\extclass}[1][]{\bandvar[#1]{extended\_class}}
\newcommand{\magauto}[1][]{\bandvar[#1]{mag\_auto}}
\newcommand{\magpsf}[1][]{\bandvar[#1]{mag\_psf}}

\newcommand{\ra}{{\ensuremath{\rm RA}}\xspace}
\newcommand{\dec}{{\ensuremath{\rm Dec}}\xspace}

\newcommand{\SExtractor}{\code{SourceExtractor}}
\newcommand{\sextractor}{\SExtractor}
\newcommand{\PSFEx}{\code{PSFEx}}

\newcommand{\SCAMP}{\code{SCAMP}}
\newcommand{\scamp}{\SCAMP}

\newcommand{\HEALPix}{\code{HEALPix}}
\newcommand{\healpix}{\HEALPix}

\newcommand{\nside}{\code{nside}}
\newcommand{\teff}{\ensuremath{t_{\rm eff}}\xspace}
\newcommand{\texp}{\ensuremath{T_{\rm exp}}\xspace}

\providecommand\physrep{\ref@jnl{Phys.~Rep.}}%
\providecommand\apjs{\ref@jnl{ApJS}}%
\providecommand{\jcap}{\ref@jnl{JCAP}}%

\newcommand{\nobjs}{2{,}500{,}247{,}752 \xspace}
\newcommand{\approxnobjs}{\ensuremath{\mathrm{2.5\ billion}}\xspace}

\newcommand{\approxnobjsgriz}{\ensuremath{\mathrm{618\ million}}\xspace}

\newcommand{\approxnstars}{\ensuremath{\mathrm{621\ million}}\xspace}

\newcommand{\approxngals}{\ensuremath{\mathrm{749\ million}}\xspace}

\newcommand{\areanimagesg}{24663} 
\newcommand{\areanimagesr}{22939} 
\newcommand{\areanimagesi}{21283} 
\newcommand{\areanimagesz}{22866} 
\newcommand{\areanimagesgriz}{16972} 

\newcommand{\approxarea}{17{,}000} 

\newcommand{\nexpg}{42034}
\newcommand{\nexpr}{41852}
\newcommand{\nexpi}{39003}
\newcommand{\nexpz}{38491}
\newcommand{\nexp}{161{,}380\xspace}
\newcommand{\approxnexp}{160{,}000}

\newcommand{\medfwhmg}{1.24} 
\newcommand{\medfwhmr}{1.10} 
\newcommand{\medfwhmi}{1.02} 
\newcommand{\medfwhmz}{1.00} 

\newcommand{\astroabs}{22} 

\newcommand{\astrorepeatg}{28}
\newcommand{\astrorepeatr}{27}
\newcommand{\astrorepeati}{28}
\newcommand{\astrorepeatz}{32}
\newcommand{\astrorepeatall}{29}

\newcommand{\photabsg}{4.4} 
\newcommand{\photabsr}{23.3} 
\newcommand{\photabsi}{7.2} 
\newcommand{\photabsz}{1.6} 

\newcommand{\photabs}{\ensuremath{20}} 

\newcommand{\photrepeatg}{4.9}
\newcommand{\photrepeatr}{5.0}
\newcommand{\photrepeati}{4.5}
\newcommand{\photrepeatz}{5.4}

\newcommand{\photgaia}{7.2}

\newcommand{\maglimpsfg}{24.3}
\newcommand{\maglimpsfr}{23.9}
\newcommand{\maglimpsfi}{23.5}
\newcommand{\maglimpsfz}{22.8}
\newcommand{\maglimpsfteng}{23.5}
\newcommand{\maglimpsftenr}{23.1}
\newcommand{\maglimpsfteni}{22.7}
\newcommand{\maglimpsftenz}{22.1}

\newcommand{\maglimautog}{23.9}
\newcommand{\maglimautor}{23.5}
\newcommand{\maglimautoi}{23.0}
\newcommand{\maglimautoz}{22.4}
\newcommand{\maglimautoteng}{22.8}
\newcommand{\maglimautotenr}{22.5}
\newcommand{\maglimautoteni}{22.1}
\newcommand{\maglimautotenz}{21.4}

\newcommand{\starefficiency}{97} 
\newcommand{\starcontamination}{2} 
\newcommand{\galefficiency}{99}
\newcommand{\galcontamination}{2}

\shorttitle{DELVE Data Release 2}

\begin{document}

\reportnum{\footnotesize FERMILAB-PUB-22-209-LDRD-PPD}
\reportnum{\footnotesize DES-2022-687}

\title{The DECam Local Volume Exploration Survey Data Release 2}


\author[0000-0001-8251-933X]{A.~Drlica-Wagner}
\affiliation{Fermi National Accelerator Laboratory, P.O.\ Box 500, Batavia, IL 60510, USA}
\affiliation{Kavli Institute for Cosmological Physics, University of Chicago, Chicago, IL 60637, USA}
\affiliation{Department of Astronomy and Astrophysics, University of Chicago, Chicago, IL 60637, USA}

\author[0000-0001-6957-1627]{P.~S.~Ferguson}
\affiliation{Physics Department, 2320 Chamberlin Hall, University of Wisconsin-Madison, 1150 University Avenue Madison, WI  53706-1390}

\author[0000-0002-6904-359X]{M.~Adam\'ow}
\affiliation{Center for Astrophysical Surveys, National Center for Supercomputing Applications, 1205 West Clark St., Urbana, IL 61801, USA}
\author{M.~Aguena}
\affiliation{Laborat\'orio Interinstitucional de e-Astronomia - LIneA, Rua Gal. Jos\'e Cristino 77, Rio de Janeiro, RJ - 20921-400, Brazil}
\author{F.~Andrade-Oliveira}
\affiliation{Department of Physics, University of Michigan, Ann Arbor, MI 48109, USA}
\author{D.~Bacon}
\affiliation{Institute of Cosmology and Gravitation, University of Portsmouth, Portsmouth, PO1 3FX, UK}
\author[0000-0001-8156-0429]{K.~Bechtol}
\affiliation{Physics Department, 2320 Chamberlin Hall, University of Wisconsin-Madison, 1150 University Avenue Madison, WI  53706-1390}
\author[0000-0002-5564-9873]{E.~F.~Bell}
\affiliation{Department of Astronomy, University of Michigan, 1085 S. University Ave., Ann Arbor, 48109-1107, USA}
\author{E.~Bertin}
\affiliation{CNRS, UMR 7095, Institut d'Astrophysique de Paris, F-75014, Paris, France}
\affiliation{Sorbonne Universit\'es, UPMC Univ Paris 06, UMR 7095, Institut d'Astrophysique de Paris, F-75014, Paris, France}
\author{P.~Bilaji}
\affiliation{Kavli Institute for Cosmological Physics, University of Chicago, Chicago, IL 60637, USA}
\affiliation{Department of Physics, University of Chicago, Chicago, IL 60637, USA}
% ...

\collaboration{124}{(DELVE Collaboration, DES Collaboration, Astro Data Lab)}

\correspondingauthor{Alex Drlica-Wagner (kadrlica@fnal.gov), Peter Ferguson (peter.ferguson@wisc.edu)}

\begin{abstract}

We present the second public data release (DR2) from the DECam Local Volume Exploration survey (DELVE). 
DELVE DR2 combines new DECam observations with archival DECam data from the Dark Energy Survey, the DECam Legacy Survey, and other DECam community programs.
DELVE DR2 consists of $\roughly \approxnexp$ exposures that cover $>21,000 \deg^2$ of the high Galactic latitude ($|b| > 10\deg$) sky in four broadband optical/near-infrared filters ($g,r,i,z$).
DELVE DR2 provides point-source and automatic aperture photometry for $\roughly \approxnobjs$ astronomical sources with a median $5\sigma$ point-source depth of $g{=}\maglimpsfg$, $r{=}\maglimpsfr$, $i{=}\maglimpsfi$, and $z{=}\maglimpsfz$ mag.
A region of $\roughly \approxarea \deg^2$ has been imaged in all four filters, providing four-band photometric measurements for $\roughly \approxnobjsgriz$ astronomical sources.
DELVE DR2 covers more than four times the area of the previous DELVE data release and contains roughly five times as many astronomical objects.
DELVE DR2 is publicly available via the NOIRLab Astro Data Lab science platform.
\end{abstract}



\section{Introduction}
\label{sec:intro}

Digital sky surveys at optical/near-infrared wavelengths have revolutionized astronomy.
These large, untargeted observational programs provide expansive data sets that enable unprecedented statistical studies and fortuitous discoveries across a wide range of astronomical fields. 
The Sloan Digital Sky Survey \citep[SDSS;][]{York:2000}, the Two Micron All-Sky Survey \citep[2MASS][]{Skrutskie:2006}, the Pan-STARRS1 survey \citep[PS1;][]{Chambers:2016}, and the SkyMapper Southern Sky Survey \citep{Wolf:2018} have provided an unprecedented view of the sky.
However, these surveys were carried out on relatively small ($\lesssim 2.5$-m diameter) telescopes, which limited their sensitivity, especially in the southern hemisphere.

The 570-megapixel Dark Energy Camera \citep[DECam;][]{Flaugher:2015} on the 4-m Victor M.\ Blanco Telescope at Cerro Tololo in Chile is the premier optical/near-infrared survey instrument in the southern hemisphere.
Since commissioning in 2012, DECam has been used by the Dark Energy Survey \citep[DES;][]{DES:2005,DES:2016}, the DECam Legacy Survey \citep[DECaLS;][]{Dey:2019}, and numerous smaller community programs.
Through these programs, DECam has gradually, and somewhat unsystematically, imaged much of the southern celestial hemisphere \citep[e.g.,][]{Nidever:2021a}.
The DECam Local Volume Exploration Survey \citep[DELVE;][]{Drlica-Wagner:2021}\footnote{\url{https://delve-survey.github.io}} seeks to complete contiguous DECam coverage of the southern sky by selectively observing regions of the sky that lack existing observations.
The primary science goals of DELVE are to discover and characterize faint satellite galaxies and other resolved stellar systems around the Milky Way, Magellanic Clouds, and isolated Magellanic analogs in the Local Volume \citep{Drlica-Wagner:2021}.
The DELVE science program has already resulted in the discovery and characterization of five ultra-faint Milky Way satellites \citep{Mau:2020,Martinez-Vazquez:2021,Cerny:2020,Cerny:2021,Cerny:2022} and an extended study of the Jet stellar stream \citep{Ferguson:2022}.
Moreover, the unprecedented wide, deep DELVE data set has broad applicability to a wide range of Galactic and extragalactic science (see \citealt{Drlica-Wagner:2021} for examples).

\begin{figure*}[t]
    \centering
    \includegraphics[width=\textwidth]{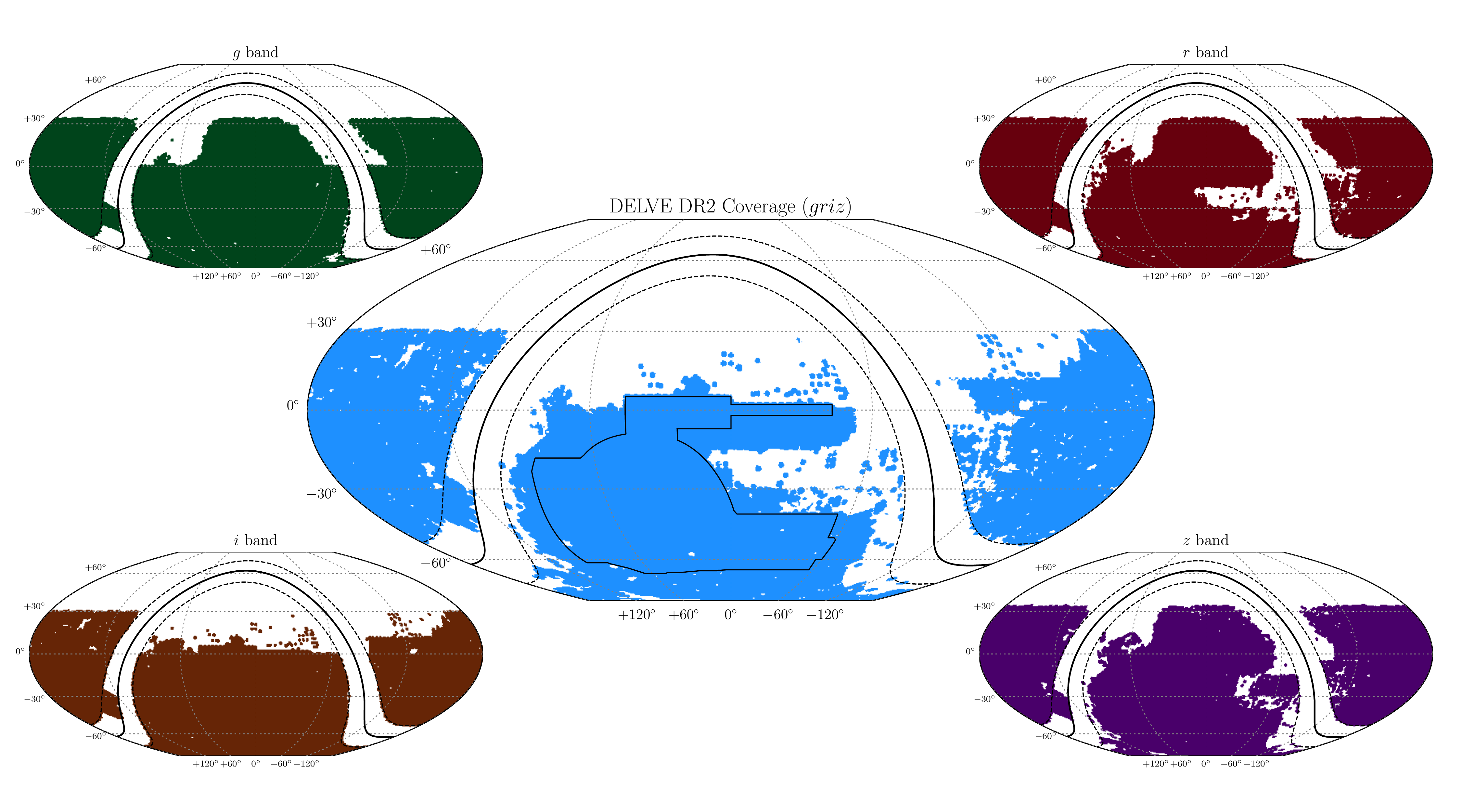}
    \caption{DELVE DR2 covers $>20{,}000 \deg^2$ in each of the $g,r,i,z$ bands (colored regions) and $\roughly \approxarea \deg^2$ in all four bands simultaneously (blue region). 
The $\roughly 5{,}000 \deg^2$ footprint of DES is outlined in black. 
These and other sky maps are shown in the equal-area McBryde-Thomas flat polar quartic projection.
}
    \label{fig:footprint}
\end{figure*}

We present the DELVE second data release (DR2), which includes imaging from DELVE, DES, DECaLS, and other public DECam programs covering ${>}21{,}000 \deg^2$ of sky in $g$, $r$, $i$, and $z$ individually and $\roughly \approxarea \deg^2$ in all four bands (\figref{footprint}).
These DECam data have been consistently processed with the DES Data Management \citep[DESDM;][]{Morganson:2018} pipeline, providing accurate point-spread function (PSF) and automatic aperture measurements for $\roughly \approxnobjs$ astronomical sources.
In this paper, we describe the DELVE DR2 data set (\secref{data}) and data reduction pipeline (\secref{processing}).
We present studies characterizing the sky coverage, astrometry, photometric calibration, depth, and object classification of the DELVE DR2 catalog in \secref{release}.
In \secref{access} we describe how the DELVE DR2 data can be accessed via the NSF's National Optical-Infrared Astronomy Research Laboratory (NOIRLab) Astro Data Lab.
Finally, we conclude in \secref{summary}.


\begin{deluxetable*}{l c c c c c}
\tablewidth{0pt}
\tabletypesize{\footnotesize}
\tablecaption{ DELVE DR2 key numbers and data quality summary. } 
\label{tab:summary}
\tablehead{
Survey Characteristic & \multicolumn{4}{c}{Band} & \colhead{Reference} \\[-0.5em]
 & $g$ & $r$ & $i$ & $z$ &
}
\startdata
Number of exposures & \nexpg & \nexpr & \nexpi & \nexpz &  \secref{data} \\
Median PSF FWHM (arcsec) & \medfwhmg & \medfwhmr & \medfwhmi & \medfwhmz &  \secref{data} \\
Sky coverage (individual bands, deg$^{2}$)  & \areanimagesg & \areanimagesr & \areanimagesi & \areanimagesz &  \secref{coverage} \\ 
Sky coverage ($g,r,i,z$ intersection, deg$^{2}$) & \multicolumn{4}{c}{\areanimagesgriz} & \secref{coverage} \\ 
Astrometric repeatability (angular distance, \mas) & \astrorepeatg & \astrorepeatr & \astrorepeati & \astrorepeatz &  \secref{astrometry} \\
Astrometric accuracy vs.\ \Gaia (angular distance, \mas) & \multicolumn{4}{c}{\astroabs} & \secref{astrometry} \\ 
Photometric repeatability (mmag)  & \photrepeatg & \photrepeatr & \photrepeati & \photrepeatz & \secref{photrel} \\
Photometric uniformity vs.\ \Gaia (mmag)  & \multicolumn{4}{c}{\photgaia} & \secref{photrel} \\
Absolute photometric uncertainty (mmag) & \multicolumn{4}{c}{$\lesssim \photabs$} & \secref{photabs} \\ 
Magnitude limit (PSF, ${\rm S/N} = 5$) & \maglimpsfg & \maglimpsfr & \maglimpsfi & \maglimpsfz & \secref{depth} \\
Magnitude limit (AUTO, ${\rm S/N} = 5$) & \maglimautog & \maglimautor & \maglimautoi & \maglimautoz & \secref{depth}\\
Galaxy selection ($\var{EXTENDED\_COADD} \geq 2$, $19 \leq \magauto[g] \leq 22$) & \multicolumn{4}{c}{Eff. $>\galefficiency\%$; Contam. $<\galcontamination\%$} & \secref{classification} \\
Stellar selection ($\var{EXTENDED\_COADD} \leq 1$, $19 \leq \magauto[g] \leq 22$) & \multicolumn{4}{c}{Eff. $>\starefficiency\%$; Contam. $<\starcontamination\%$} & \secref{classification} \\
\enddata
\end{deluxetable*}

\section{Data Set}
\label{sec:data}

DELVE DR2 is comprised of \nexp DECam exposures assembled from ${>}\,270$ DECam community programs (\appref{propid}).
The largest contributors to the DELVE DR2 data set are DES \citep{DES-DR2:2021}, DECaLS \citep{Dey:2019}, DELVE \citep{Drlica-Wagner:2021}, and the DECam eROSITA Survey (DeROSITAS; PI Zenteno)\footnote{\url{http://astro.userena.cl/derositas}}.
DELVE DR2 more than quadruples the sky area of DELVE DR1 by including exposures in the southern Galactic cap ($b < -10\degree$) and exposures in the northern celestial hemisphere (${\rm Dec.} > 0 \degree$). 
In addition, DELVE and DeROSITAS have continued to observe regions of the sky that lack DECam imaging to increase the coverage and uniformity of the DECam data set \citep[see Section 3 of][]{Drlica-Wagner:2021}.
The key properties of the DELVE DR2 data set are listed in \tabref{summary}.

Separate criteria were used to select input exposures in the northern Galactic cap, the southern Galactic cap, and the DES region.
The northern Galactic cap data set is comprised of DECam exposures with $b > 10\degree$ plus an extension into the Galactic plane ($b > 0\degree$) in the region of $120\degree < \ra < 140\degree$ to enable an extended analysis of the Jet stellar stream \citep{Jethwa:2018,Ferguson:2022}.
Exposures in the southern Galactic cap were selected to have $b < -10 \degree$, excluding exposures within the DES footprint and exposures collected by the DES program. 
The DES exposures reside in the southern Galactic cap, but they were selected separately when defining the input to DES DR2 \citep{DES-DR2:2021}.

\begin{figure*}[t]
    \centering
    \includegraphics[width=0.49\textwidth]{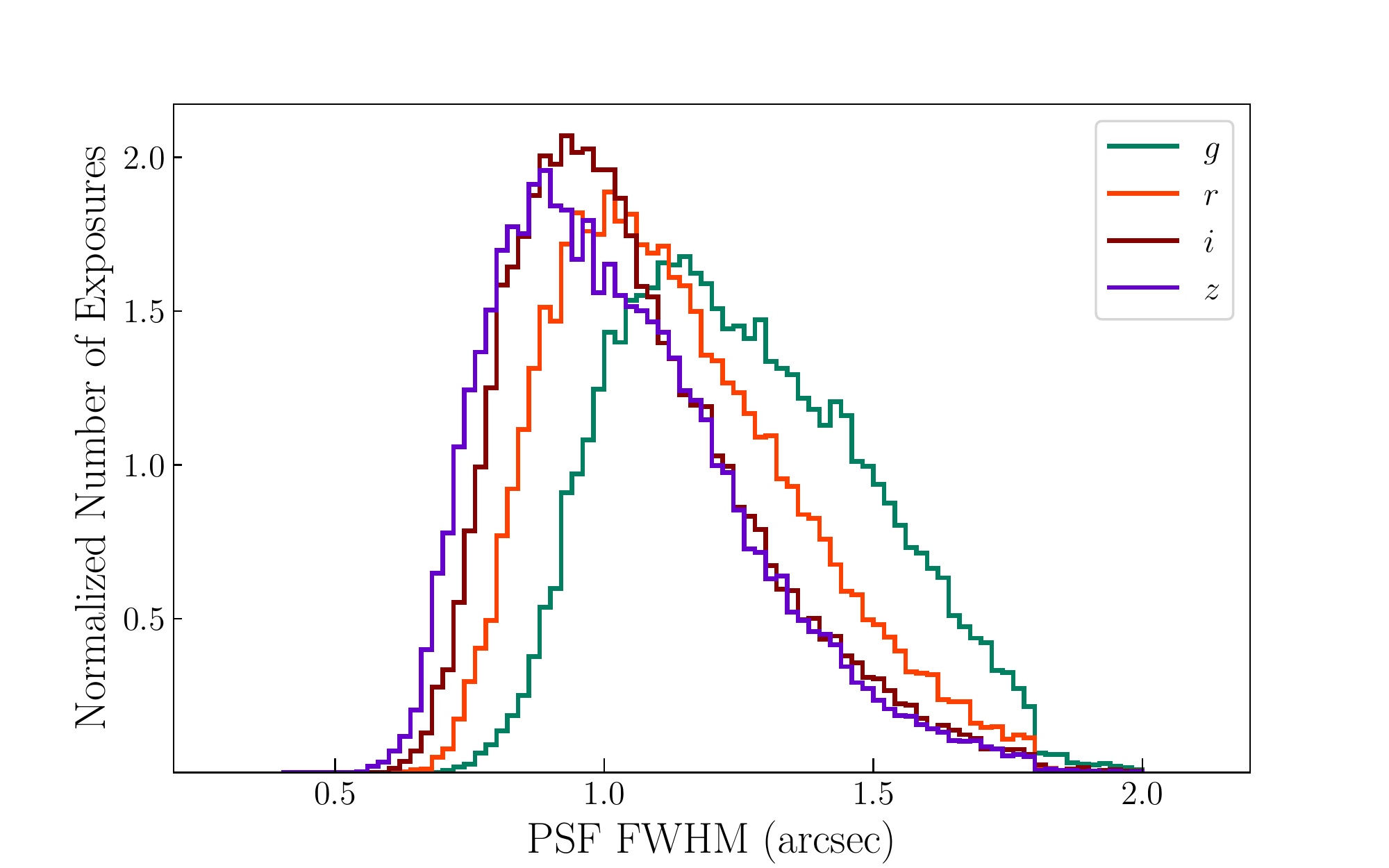}
    \includegraphics[width=0.49\textwidth]{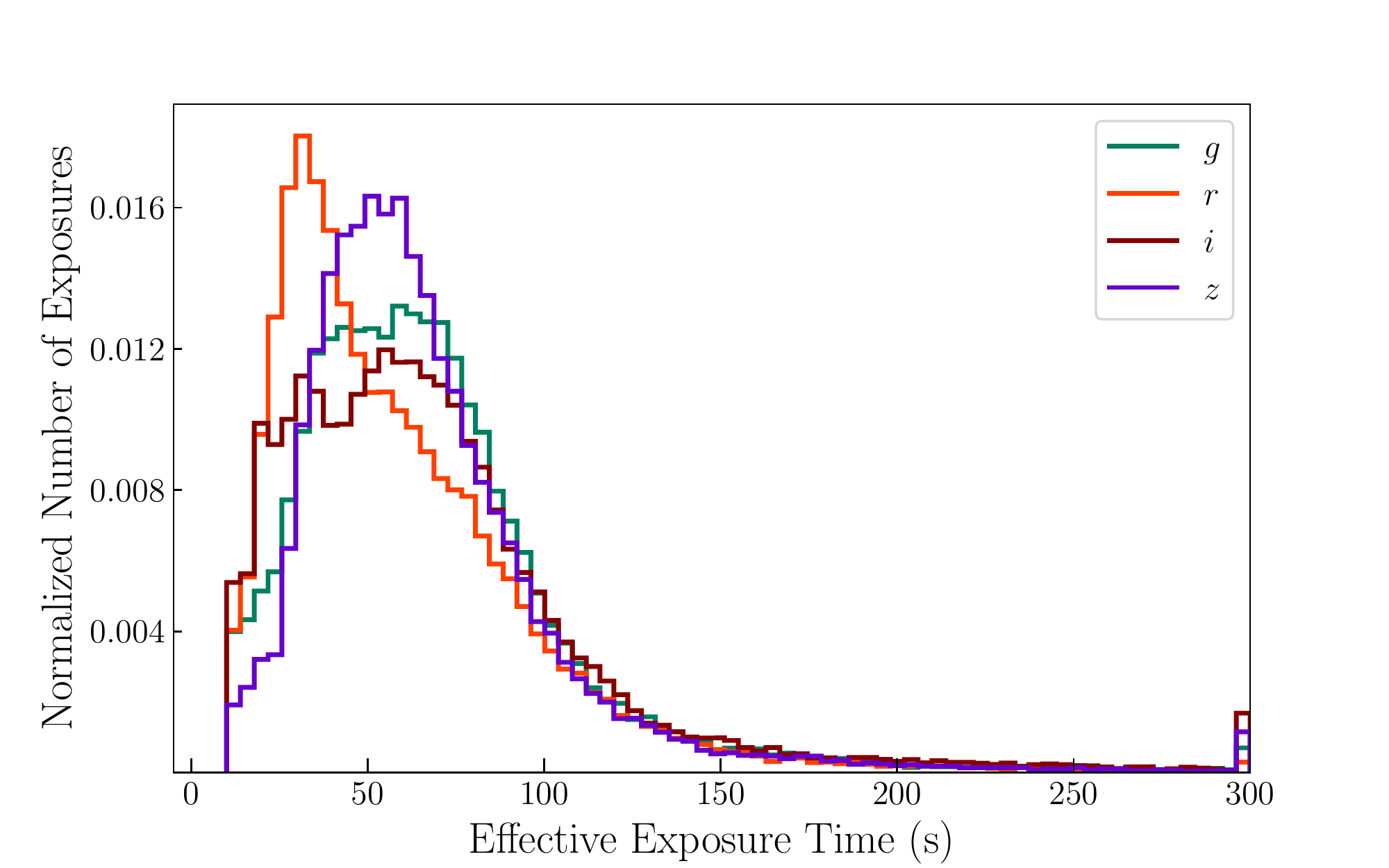}
    \caption{(Left) PSF FWHM distributions for DECam exposures included in DELVE DR2. (Right) Distributions of effective exposure time ($\teff \times \texp$) for exposures included in DELVE DR2.}
    \label{fig:fwhm}
\end{figure*}

For each exposure, we calculate the effective depth based on the effective exposure time scale factor, \teff, which compares the achieved seeing, sky brightness, and extinction due to clouds relative to canonical values for the site \citep{Neilsen:2016}.
Exposures in the northern Galactic cap region were required to have an effective exposure time scale factor of $\teff > 0.3$.
The requirement on $\teff$ was relaxed in the southern Galactic cap to avoid rejecting exposures taken close to the southern celestial pole.
These exposures are observed at high airmass ($\sec(z) \sim 2$) and have systematically worse PSF full width at half maximum (FWHM).
Exposures in the southern Galactic cap were required to have $\teff > 0.2$ and $\teff \times \texp > 12\second$.
No explicit cut was placed on the PSF FWHM in the northern Galactic cap (the cut on \teff removes exposures with very poor seeing), while a cut of ${\rm FWHM} < 1\farcs8$ was applied in the southern Galactic cap. 
The resulting distribution of PSF FWHM and effective exposure time for the full DELVE DR2 data set are shown in \figref{fwhm}.

All exposures in the northern and southern Galactic caps were required to have good astrometric solutions when matched to \Gaia DR2 \citep{Gaia:2018b} by \SCAMP \citep{Bertin:2006}.
These criteria required $>250$ astrometric matches, $\chi^2_{\rm astrom} < 500$, $\Delta(\ra) < 150\mas$, and $\Delta(\dec) < 150\mas$.
We identified and removed exposures that were heavily contaminated by spurious scattered and reflected light from bright stars using the ray-tracing procedure developed by DES \citep{Kent:2013}.
In addition, rare failures in the sky background estimation can cause a large number of spurious object detections. 
A handful of exposures suffering from this processing failure were identified as having a large fraction of unmatched objects, and they were removed from the final catalog production.

DELVE DR2 includes $\roughly 60,000$ exposures collected by DES that were processed and calibrated as input into DES DR2 \citep{DES-DR2:2021}.\footnote{DELVE DR2 does not include the DES $Y$-band imaging.} 
The DES processing pipeline required $\teff > 0.2$ for $g$-band exposures and $\teff > 0.3$ for exposures taken in $r$, $i$, and $z$.
DES applied a wavelength-dependent criterion to remove exposures with poor PSF FWHM resulting in a maximum PSF FWHM of $\{ 1\farcs72, 1\farcs62, 1\farcs56, 1\farcs50\}$ in $g,r,i,z$, respectively.
Additional cuts were applied to remove exposures that were contaminated by stray or scattered light, airplanes, excessive electronic noise, and other artifacts.
A full description of the DES data selection and processing criteria can be found elsewhere \citep{Morganson:2018,DES-DR1:2018,DES-DR2:2021}.

\section{Data Processing}
\label{sec:processing}

All exposures in DELVE DR2 were processed with the DESDM ``Final Cut'' pipeline \citep{Morganson:2018} as implemented for the processing of DES DR2 \citep{DES-DR2:2021}.
Data were reduced and detrended using seasonally averaged bias and flat images, and full-exposure sky background subtraction was performed \citep{Bernstein:2018}. 
\SExtractor \citep{Bertin:1996} and \PSFEx \citep{Bertin:2011} were used to automate source detection and photometric measurement. 
Astrometric calibration was performed against \Gaia DR2 using \SCAMP \citep{Bertin:2006}.\footnote{Associated configuration files can be found at: \url{https://github.com/delve-survey/delve_config}.}
We note that DELVE DR2 does not include the production of coadded images \citep[e.g.,][]{DES-DR1:2018,DES-DR2:2021}; however, we expect that coadded images will be produced as part of a future DELVE data release.

Photometric zeropoints for each DECam CCD were derived independently for the DES exposures and the other DECam exposures included in DELVE DR2.
For the DES exposures, we applied zeropoints that were derived for DES DR2 using the forward global calibration module \citep[FGCM;][]{Burke:2018}. 
The FGCM procedure fits time-dependent atmospheric and instrumental conditions to establish an internal network of calibration stars.
These calibration stars are then used to iteratively refine the photometric calibration of exposures taken during both photometric and non-photometric conditions.
The FGCM has been demonstrated to achieve a relative photometric calibration uncertainty of $\roughly 2 \mmag$ when applied to the DES exposures \citep{DES-DR2:2021}.
In contrast, the non-DES exposures included in DELVE DR2 were calibrated following the simple external calibration procedure developed for DELVE DR1 \citep{Drlica-Wagner:2021}.
Briefly, we performed a $1\arcsec$ match between objects in the Final Cut catalogs for each DECam CCD and the ATLAS Refcat2 catalog \citep{Tonry:2018}.
ATLAS Refcat2 covers the entire sky by placing measurements from PS1 DR1 \citep{Chambers:2016}, SkyMapper DR1 \citep{Wolf:2018}, and several other surveys onto the PS1 $g,r,i,z$-bandpass system.
Transformation equations from the ATLAS Refcat2 system to the DECam system were derived by comparing calibrated stars from DES DR1 (Appendix A of \citealt{Drlica-Wagner:2021}).
Zeropoints were derived by finding the median offset required to match the DECam observations to the matched ATLAS Refcat2 observations.
Zeropoints derived from the DELVE processing and photometric calibration pipeline were found to agree with those derived by DES DR2 with a scatter of $\roughly 10 \mmag$.
While the external calibration against ATLAS Refcat2 yields a significantly larger scatter than the FGCM, it can be quickly and easily applied to any DECam exposure.

We built a multi-band catalog of unique sources by combining the \SExtractor catalogs from each individual CCD image following the procedure described in \citet{Drlica-Wagner:2021}.
We took the set of \SExtractor detections with $\var{flags} < 4$, which allowed neighboring and deblended sources, and $(\var{imaflags\_iso}\,\&\,2047) = 0$, which removed objects containing bad pixels within their isophotal radii \citep{Morganson:2018}.
We further required each detection to have a measured automatic aperture flux, a measured PSF flux, and a PSF magnitude error of $< 0.5$ mag.
We sorted \SExtractor detections into $\roughly 3 \deg^2$ ($\nside=32$) \healpix pixels \citep{Gorski:2005}, and within each \healpix pixel we grouped detections into clusters by associating all detections within a $0\farcs5$ radius.
This matching radius was chosen to be significantly larger than the astrometric uncertainty (\secref{astrometry}), but smaller than the PSF FWHM (\figref{fwhm}).
Furthermore, we identified and split pairs of closely separated objects that were observed in the same image \citep{Drlica-Wagner:2021}.

Each cluster of detections was associated with an object in the DELVE DR2 catalog.
The astrometric position of each object was calculated as the median of the individual single-epoch measurements of the object.
We track two sets of photometric quantities for each object: (1) measurements from the single exposure in each band that has the largest effective exposure time (i.e., the largest $\teff \times T_{\rm exp}$), and (2) the weighted average of the individual single-epoch measurements (these quantities are prefixed by \var{WAVG}).
The weighted average and unbiased weighted standard deviation were calculated following the weighted sample prescriptions used by DES (Appendix A of \citealt{DES-DR2:2021}).\footnote{Note that we do not apply the ``error floor'' applied by DES.}
In addition, we track cluster-level statistics such as the number of detections in each band.

We follow the DES procedure to calculate the interstellar extinction from Milky Way foreground dust \citep{DES-DR1:2018}.
We compute the value of $E(B-V)$ at the location of each catalog source by performing a bi-linear interpolation in $(\ra,\dec)$ to the maps of \citet{Schlegel:1998}.
The reddening correction for each source in each band, $A_b = R_b \times E(B-V)$, is calculated using the fiducial interstellar extinction coefficients from DES DR1 \citep{DES-DR1:2018}: $R_g = 3.185$, $R_r = 2.140$, $R_i = 1.571$, and $R_z = 1.196$.
Note that, following the procedure of DES DR1, the \citet{Schlafly:2011} calibration adjustment to the \citet{Schlegel:1998} maps is included in our fiducial reddening coefficients ($N=0.78$).
The $A_b$ values are included for each object in DELVE DR2, but they are not applied to the magnitude columns by default.
The list of the photometric and astrometric properties provided in DELVE DR2 can be found in \appref{tables}.


\subsection{Improvements Relative to DELVE DR1}
\label{sec:updates}

We have made several improvements to the pipeline described by \citet{Drlica-Wagner:2021}.

\begin{enumerate}
    \item The seasonally averaged bias and flat images used for image detrending have been updated to include calibration products from the fifth and sixth years of DES observing. The final epoch of DES calibration products have been used to process all exposures taken after the end of DES data taking.
    \item Images that were heavily affected by reflected or scattered light from bright stars were identified using the DES ray-tracing tool \citep{Kent:2013}. Objects detected on these CCDs were removed from the DELVE DR2 catalog.
    \item The radius for matching sources within and across bands has been reduced from $1\arcsec$ to $0\farcs5$. This change was motivated by the excellent astrometric precision of the DELVE DR1 catalog ($\roughly 30 \mas$). The change, along with improvements in the algorithm for splitting pairs of closely separated objects, reduces the number of objects that are spuriously merged.
\end{enumerate}


\section{Data Release}
\label{sec:release}

DELVE DR2 is derived from DECam data covering $> 20{,}000 \deg^2$ in each of the $g,r,i,z$ bands, while $\roughly \approxarea \deg^2$ are jointly covered in all four bands (\figref{footprint}).
DELVE DR2 consists of a catalog of $\roughly \approxnobjs$ unique astronomical objects, with $\roughly \approxnobjsgriz$ objects that have measurements in all four bands.
This section describes the characterization of the sky coverage, astrometry, photometry, depth, and object classification of the DELVE DR2 catalog.
Summary statistics of this characterization are given   in \tabref{summary}. 

\subsection{Sky Coverage}
\label{sec:coverage}

We quantify the area covered by DELVE DR2 by pixelizing the geometry of each DECam CCD using the \code{decasu}\footnote{\url{https://github.com/erykoff/decasu}} package built on \code{healsparse}.\footnote{\url{https://healsparse.readthedocs.io}}
This package maps the geometry of each CCD using higher-resolution nested \healpix maps ($\nside = 16384; \roughly 166 \asec^2$) and sums the resulting covered pixels to generate lower resolution maps ($\nside=4096; \roughly 0.74 \amin^2$) containing the fraction of each pixel that is covered by the survey.
We quantitatively estimate the covered area as the sum of the coverage fraction maps in each band independently and the intersection of the maps in all four bands (\tabref{summary}).
 

\begin{figure*}[t!]
    \centering
    \includegraphics[width=0.49\textwidth]{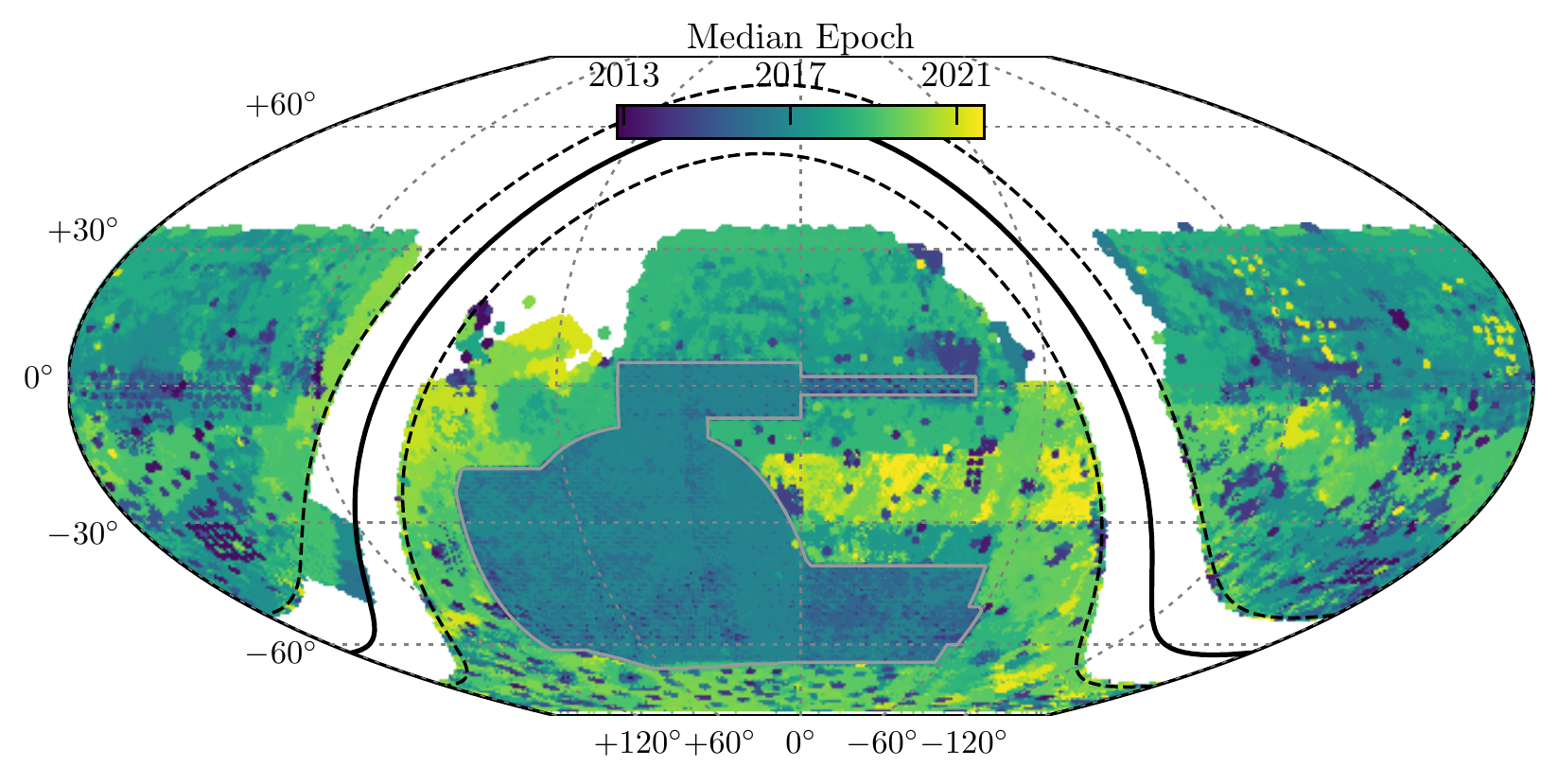}
    \includegraphics[width=0.49\textwidth]{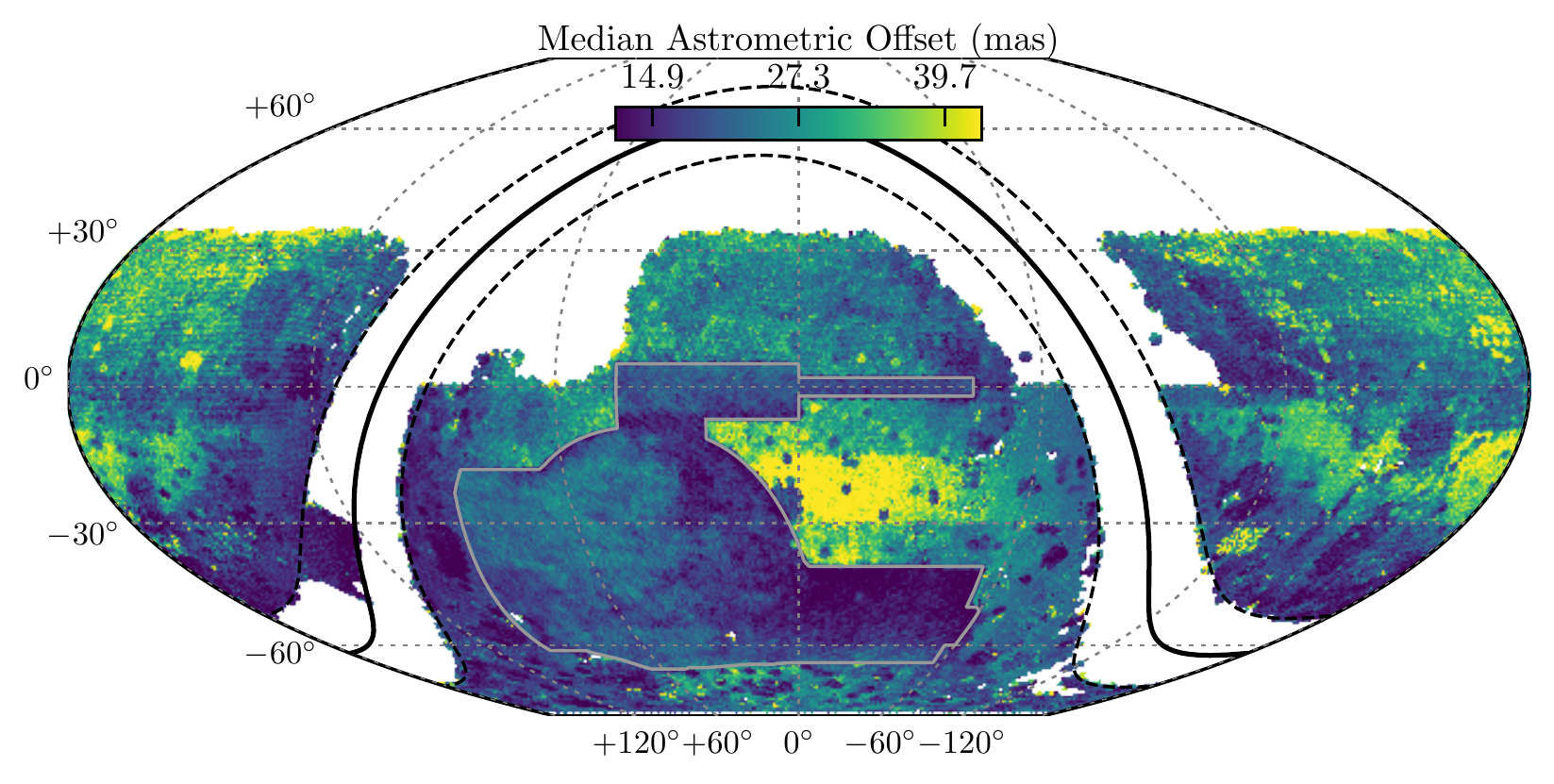}
    \caption{\label{fig:mjd_astrom} \textit{Left}: Median observational epoch for DECam observations in all bands ($griz$) that are used for calculating the coordinates of DELVE DR2 objects. \textit{Right}: Median astrometric offsets between DELVE DR2 objects with $16 < g < 19$ and \Gaia EDR3 objects matched within $2\arcsec$. Note that no correction has been made for the proper motions of objects.}
\end{figure*}

\subsection{Astrometry}
\label{sec:astrometry}

We assess the internal astrometric repeatability by comparing the distributions of angular separations of individual detections of the same objects over multiple exposures.
The median global astrometric spread is $\astrorepeatall \mas$ across all bands and is found to be fairly consistent within each band (\tabref{summary}).
Furthermore, we estimate the external astrometric accuracy by calculating the angular separation between bright stars in DELVE DR2 ($16 < g < 19$) and sources in \Gaia EDR3 \citep{Gaia:2021} matched within $2\arcsec$ (\figref{mjd_astrom}). 
We find that the median separation between the positions measured by DELVE DR2 and \Gaia EDR3 is $\astroabs \mas$, which confirms that no significant astrometric offsets have been introduced by the catalog coaddition procedure.
Since the DESDM astrometric calibration does not incorporate proper motions, we expect some correlation between the astrometric residuals and the median measurement epoch of each source (\figref{mjd_astrom}).


\begin{figure*}
    \centering
    \includegraphics[width=0.98\textwidth]{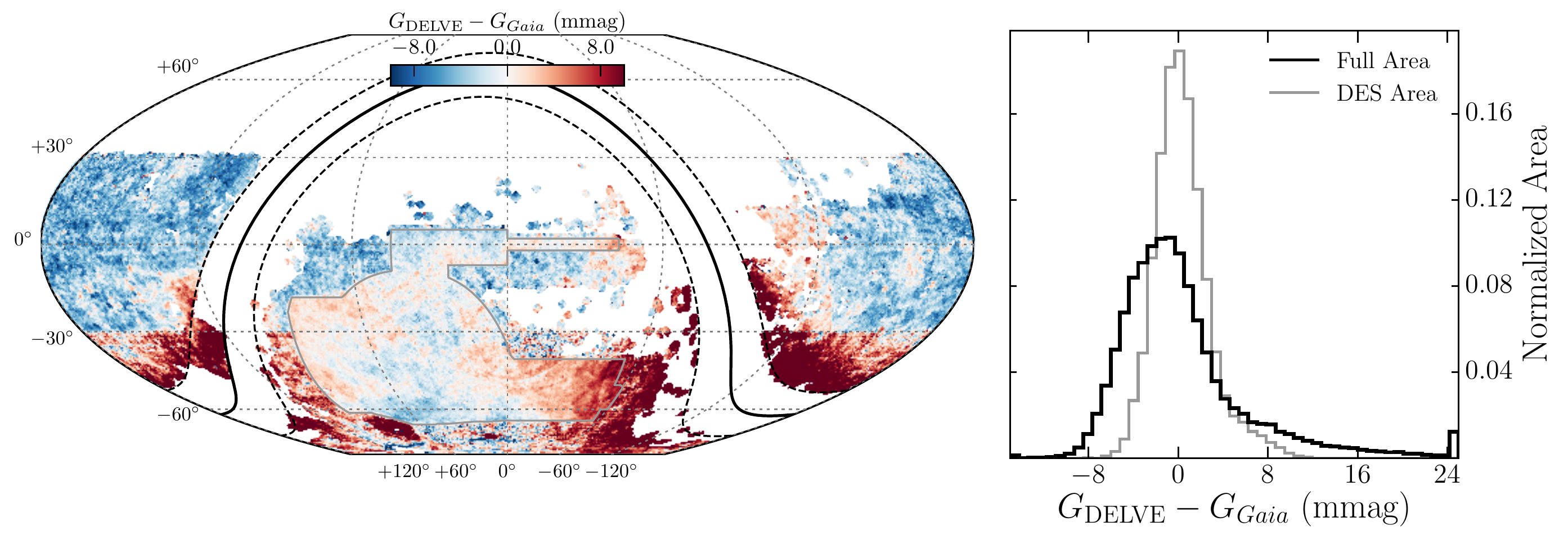}
     \caption{Median difference between the DELVE DR2 photometry transformed into the \Gaia $G$-band, $G_{\rm DELVE}$, and the measured magnitude from \Gaia EDR3, $G_\Gaia$. The spatial distribution of the median difference in each pixel is shown in the left panel (color range clipped to $\pm 10\mmag$), while the right panel shows a histogram of the pixel values. A shift in the zeropoint can be seen at $\dec. \sim -30\deg$, which corresponds to the boundary between the ATLAS Refcat2 use of PS1 and SkyMapper (\secref{photrel}).
This comparison is restricted to the area with overlapping DELVE DR2 coverage in all four bands ($g,r,i,z$).
}
    \label{fig:gaia}
\end{figure*}

\subsection{Relative Photometric Calibration}
\label{sec:photrel}

We assess the photometric repeatability in each band from the root-mean-square (rms) scatter between independent PSF magnitude measurements of bright stars. 
For each band, we select stars with $16 < \var{WAVG\_MAG\_PSF} < 18$ mag and calculate the median rms scatter in $\roughly 0.2 \deg^2$ \healpix pixels ($\nside=128$).
We estimate the median of the rms scatter over the entire footprint in each band.
This quantity is found to be $\roughly 5 \mmag$ and is listed for each band in \tabref{summary}.

We validate the photometric uniformity of DELVE DR2 by comparing to space-based photometry from \Gaia EDR3 (\figref{gaia}).
We transform the $g,r,i,z$ photometry from DELVE to the \Gaia $G$ band using a set of transformations derived for DES DR2 \citep{Sevilla-Noarbe:2020,DES-DR2:2021}.
We compare the \Gaia EDR3 $G$-band magnitude in the AB system ($G_\Gaia$) to the predicted $G$-band magnitude of stars in DELVE ($G_{\rm DELVE}$).
We calculate the median difference, $G_{\rm DELVE} - G_\Gaia$, within each $\nside=128$ \healpix pixel for stars with $16 < r < 20 \magn$, $0.5 < (g-i) < 1.5 \magn$, and \Gaia $G < 20\magn$.
We plot the spatial distribution of the median difference along with histograms for the median difference within the DES region and over the full DELVE DR2 footprint in \figref{gaia}.
While the median difference within the DES footprint is zero by construction, we find a small ($<1 \mmag$) offset between DELVE DR2 and \Gaia EDR3. 
We estimate the photometric uniformity of DELVE DR2 as the standard deviation of the median differences across pixels, which yields a value of \photgaia \mmag (\tabref{summary}).
However, because the distribution of residuals is non-Gaussian (\figref{gaia}), we also provide the 68\% containment interval, which is $9.1 \mmag$.
We find no significant magnitude-dependent trends in $G_{\rm DELVE} - G_\Gaia$ within the magnitude range that we study ($16 < r < 20 \magn$).

Similar comparisons between DES DR2 and \Gaia DR2 demonstrated that the nonuniformity of \Gaia observations can be the dominant contributor to photometric nonuniformity estimated using this technique \citep{Burke:2018,Sevilla-Noarbe:2020,DES-DR2:2021}.
Within the DES footprint, we find that comparing to \Gaia EDR3 reveals much less structure than was seen when comparing to \Gaia DR2 \citep{DES-DR2:2021}.
Furthermore, it is clear that outside the DES footprint spatial structure in the DELVE DR2 calibration dominate the nonuniformity relative to \Gaia.
We observe a systematic shift of $\roughly 10 \mmag$ relative to \Gaia EDR3 at $\dec = -30\deg$ where ATLAS Refcat2 switches from using PS1 to SkyMapper \citep{Tonry:2018,Drlica-Wagner:2021}. 
It should be possible to improve the relative photometric calibration of DELVE by applying the FGCM \citep{Burke:2018}.
Initial tests using several thousand square degrees of the DELVE data suggest that a relative photometric uniformity of $\lesssim 5\mmag$ is possible.

\begin{figure*}
    \centering
    \includegraphics[width=0.98\textwidth]{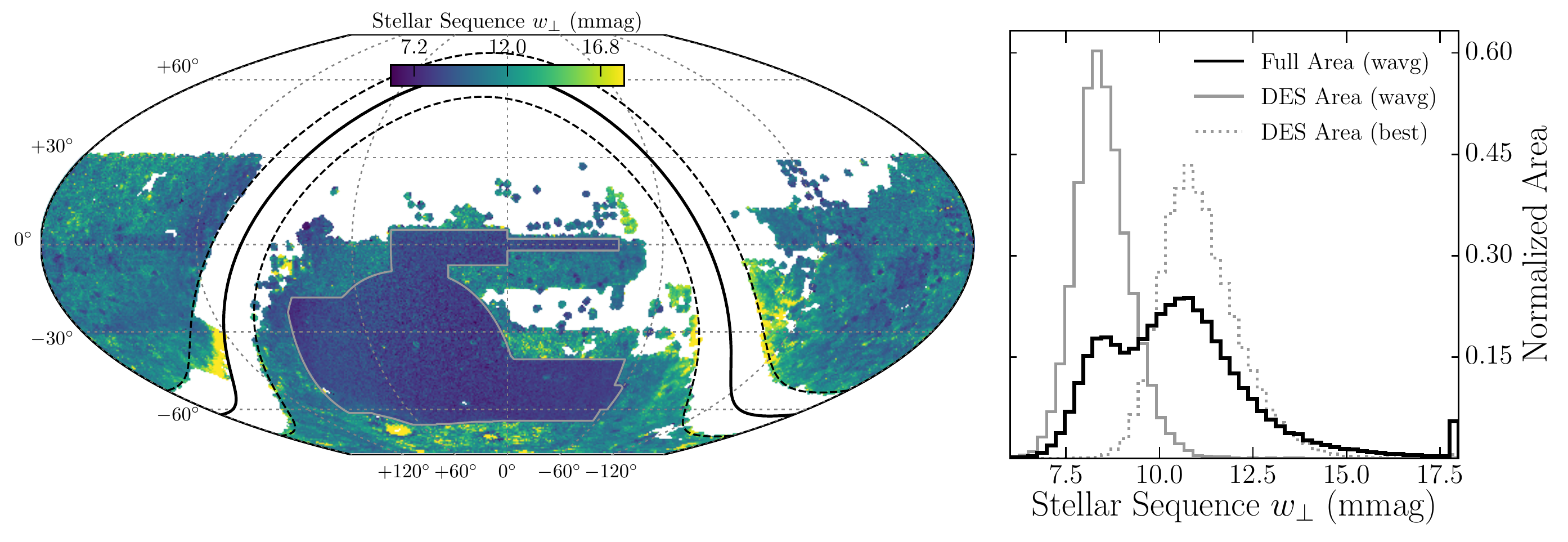}
    \caption{\emph{Left:} Spatial distribution of the measured width of the stellar locus $w_{\perp}$ using \code{WAVG\_MAG\_PSF} magnitudes for each \code{nside}=128 \healpix pixel in the DELVE DR2 footprint. The DES region can be seen to have much smaller values of $w_{\perp}$ indicating lower statistical error in these measurements. 
    \emph{Right:} Histogram of $w_{\perp}$ values ($w_\perp=\sqrt{\sigma^2+w_{\perp,0}^2}$), where $w_{\perp,0} \sim 8$ mmag. 
    The black line shows the same data as the spatial map (\code{WAVG\_MAG\_PSF} magnitudes for the full footprint, a clear bi-modality can be seen due to the difference in relative statistical error in measurements between the DES region calibrated with FGCM ($\sigma_{(\mathrm{FGCM},\,\code{WAVG})} \sim 3$ mmag), and the rest of the DELVE footprint calibrated with ATLAS Refcat2 ($\sigma_{(\mathrm{ATLAS\, R2},\,\code{WAVG})} \sim 7$ mmag).
    The gray histograms illustrate the difference in the measured width between the weighted-average (solid) and single best measurements (dotted).}
    \label{fig:wperp}
\end{figure*}

\begin{figure}
    \centering
    \includegraphics[width=0.98\columnwidth]{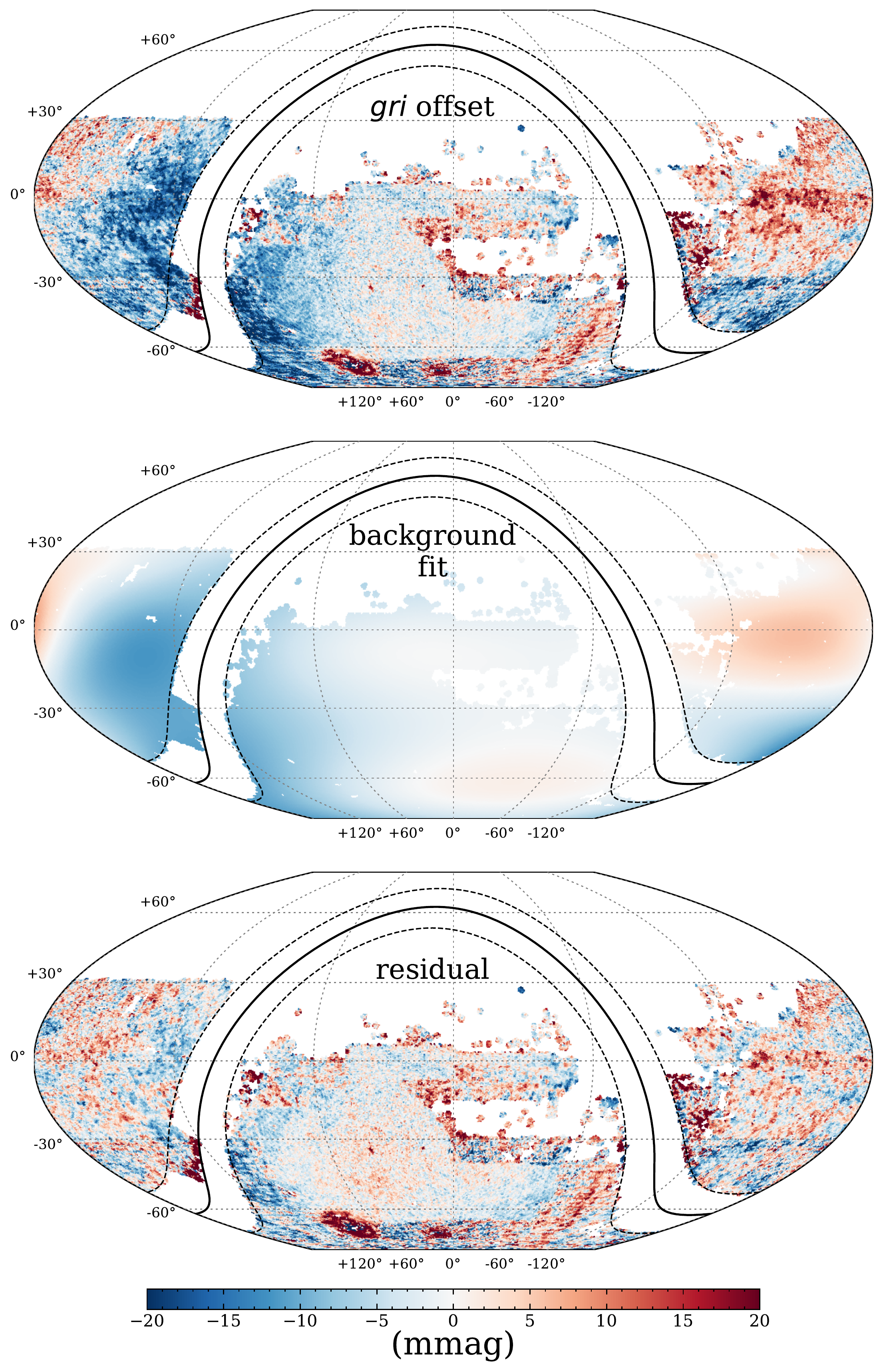}
    \caption{\emph{Top:} Offset in the stellar locus $(r-i)$ color at $(g-r)=0.7$ fit in each \code{nside}=128 \healpix pixel relative to the DES value of $(r-i)_{\rm DES}=0.221$ mag. 
Offsets in this distribution at large spatial scales are likely due to changing stellar populations.  
    \emph{Middle:} Polynomial fit to the $(r-i)$ offset map smoothed with a $\sigma=5\degree$ Gaussian kernel. 
    \emph{Bottom:} Map of residuals after the polynomial fit has been subtracted. 
This residual map highlights variations in the location of the stellar locus at smaller scales and is an estimate of the color uniformity.}
    \label{fig:slr}
\end{figure}

\subsection{Color Uniformity} 
\label{sec:slr}
As an additional check of the color uniformity and relative photometric calibration of DELVE DR2, we perform an analysis of the stellar sequence using the $g$, $r$, and $i$ bands \citep[e.g.,][]{Ivezic:2004a,MacDonald:2004a,High:2009a,Gilbank:2011a,Coupon:2012a,Kelly:2014,Drlica-Wagner:2018}.
The stellar sequence follows a tight locus in the $(g-r)$ vs.\ $(r-i)$ color-color plane, especially in the region from $0.3 < (g-r) < 1.1$. 
This region of the stellar sequence is dominated by main sequence stars and has a small intrinsic width. 
This tight relation allows us to assess the calibration quality in two ways:
(1) On small scales, we can probe the statistical error in color measurements by computing the width of the stellar sequence ($w_\perp$). 
(2) On larger angular scales, we can use variations in the location of this sequence as an estimate of systematic color uniformity.

We follow the methodology of \citet{Ivezic:2004a} to measure both the width and location of the stellar sequence.
Briefly, we select high confidence stars ($\var{EXTENDED\_CLASS\_G}=0$) that are bright with $g$, $r$, and $i$ extinction-corrected magnitudes brighter than $20\magn$ and extinction-corrected color $0.3 < (g-r) < 1.1$.

We performed a linear fit on the data and derived principal components, $P_1$ and $P_2$, where $P_2$ is perpendicular to the stellar locus line of best fit. 
 
We define $w_\perp$ to be the $3\sigma$-clipped rms of the distribution of stars in the $P_2$ direction. 
The location of the stellar sequence is summarized as a residual between the $(r-i)$ color of the linear fit at $(g-r)=0.7$.
This value is computed relative to a low extinction ($E(B-V) < 0.015$) empirical stellar locus computed from the DES DR2 catalog, where $(r-i)_{\rm DES}=0.221\magn$ at $(g-r)_{\rm DES}=0.7 \magn$.

To estimate the magnitude of the statistical error on color we split our data set into two areas. 
First, we analyze the DES footprint, which is covered homogeneously and has zeropoints derived from FGCM. 
Second, we analyze the rest of the DELVE DR2 footprint where zeropoints were derived from ATLAS Refcat2 (\secref{processing}). 
We calculate the width of the stellar sequence, $w_\perp$, using both the best single-epoch measurement (\code{MAG\_PSF}) and the weighted-average catalog coadd measurements (\code{WAVG\_MAG\_PSF}) for each $\nside = 128$ \healpix pixel. 
The spatial distribution of $w_\perp$ derived from the weighted-average magnitudes can be seen in \figref{wperp}.
For the region in the DES footprint, we also compute an estimate of the relative difference in the statistical errors between each type of magnitude measurement, $N_{eff}=\code{MAGERR\_PSF}^2/\code{WAVG\_MAGERR\_PSF}^2$. 
Assuming that $w_\perp$ comes from the statistical uncertainty in the photometric calibration ($\sigma_{\rm stat}$) and intrinsic width of the stellar sequence ($w_{\perp,0}$) added in quadrature ($w_\perp^2=\sigma_{\rm stat}^2+w_{\perp,0}^2$), we can use the two measurements of $w_{\perp}$ and effective number of observations ($N_{eff}$) for the \code{WAVG} measurement to solve for $\sigma_{stat}$ and $w_{\perp,0}$. 

Distributions for $w_\perp$ in the DES region for the single measurement and \code{WAVG} measurement cases are shown on the right of \figref{wperp} in gray. 
We find a median single measurement (\code{WAVG} measurement) error of $\sigma_{(\mathrm{FGCM})} \sim 8$ mmag ($\sigma_{(\mathrm{FGCM},\,\code{WAVG})} \sim 3$ mmag) for the region with zeropoints derived from FGCM, and median intrinsic width of the stellar locus $w_{\perp,0} \sim 8$ mmag.
To estimate $\sigma_{\rm stat}$ for the ATLAS Refcat2 calibrated region where the coverage is not as homogeneous, we use the $w_{\perp,0}$ estimate from the FGCM region. 
The median single measurement (\code{WAVG} measurement) error of $\sigma_{(\mathrm{ATLAS\, R2})} \sim 10$ mmag ($\sigma_{(\mathrm{ATLAS\, R2},\,\code{WAVG})} \sim 7$ mmag) for the region with zeropoints derived from ATLAS Refcat2. 
This value of $\sigma_{(\mathrm{ATLAS\, R2},\,\code{WAVG})}$ agrees with the comparison to \Gaia EDR3 data in \secref{photrel}. 
Furthermore, this analysis highlights the differences in color uncertainty between the FGCM calibrated region and the ATLAS Refcat2 calibrated region. 
We note that variations in reddening and underlying stellar populations could cause variations in the intrinsic width of the stellar locus, and our value in the DES region of $w_{\perp,0}= 8$ mmag can be thought of as a lower limit over the rest of the sky.
Therefore, the inferred $\sigma_{(\mathrm{ATLAS\, R2})}$ is an upper limit on the statistical color uncertainty in the ATLAS Refcat2 calibrated region.   

As described above, we use the position of the stellar locus in the $(g-r)$ vs.\ $(r-i)$ plane as a probe of color uniformity in DELVE. Similar to $w_{\perp}$, we use the results of our fit calculated for each $\nside = 128$ \healpix pixel. 
The offsets between the calculated value and the DES Y6 value for each \healpix pixel are shown in the top of Figure \ref{fig:slr}. 
Using \code{MAG\_PSF} (\code{WAVG\_MAG\_PSF}) we find a median rms in the $(r-i)$ color of the linear fit at $(g-r)=0.7$ of 9 mmag (8 mmag) for the entire survey footprint, with a scatter between \code{MAG\_PSF} and \code{WAVG\_MAG\_PSF} of less than 3 mmag. 
If we compare the DES footprint to the rest of the DELVE using \code{MAG\_PSF}, we find median rms measurements of 5 mmag and 9 mmag respectively. 
It is likely that some of this scatter can be attributed to the effects of interstellar extinction and changes in the observed stellar populations across the footprint, which will shift the location of the stellar locus \citep[see Section 2.3 of][]{High:2009a}. 
To estimate the effect of reddening on these values, we compute a median rms only for regions with $E(B-V) < 0.5$\,mag and find that our results are unchanged. 
This indicates that reddening systematics do not strongly contribute to the spatial structure seen in the top row of Figure \ref{fig:slr}.
In order to account for shifts of the stellar locus on large spatial scales (tens of degrees) and estimate the color uniformity on scales of a few degrees, we smooth the spatial distribution of the residuals with a Gaussian kernel with a standard deviation of $\sigma = 5\degree$ and fit a 5th order polynomial. 
This polynomial is then subtracted from the spatial distribution, mitigating the effect of spatially dependent changes in the location of the stellar locus and highlighting systematic scatter in the color uniformity at scales of a few degrees. 
Using this subtracted map, we find a median rms of 4\,mmag for the DES region and 7\,mmag for the rest of the DELVE DR2 footprint.
This can be interpreted as a lower limit on the systematic uncertainties in the color measurements of DELVE DR2.

\begin{figure*}[t]
    \centering
    \includegraphics[width=0.49\textwidth]{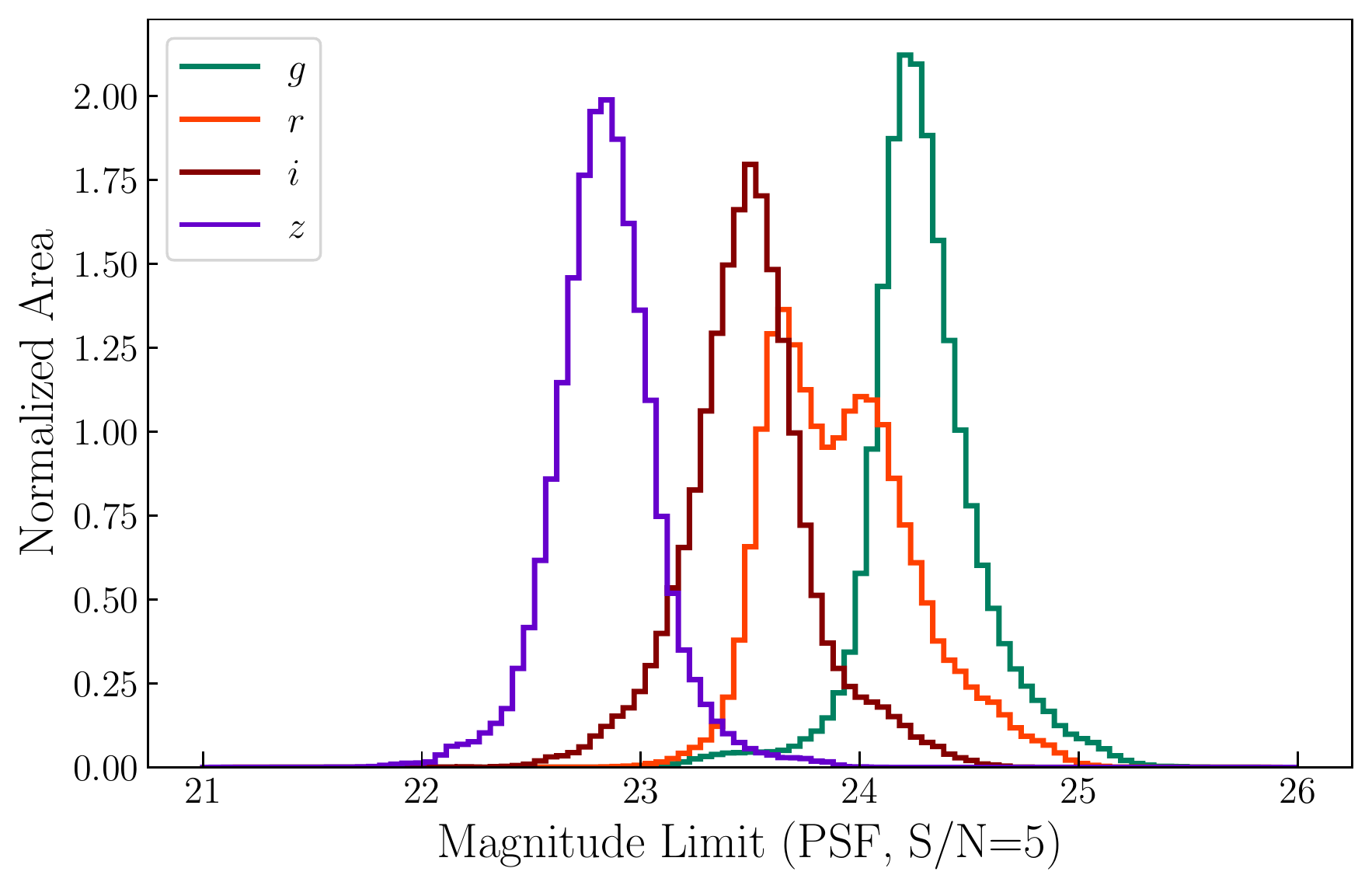}
    \includegraphics[width=0.49\textwidth]{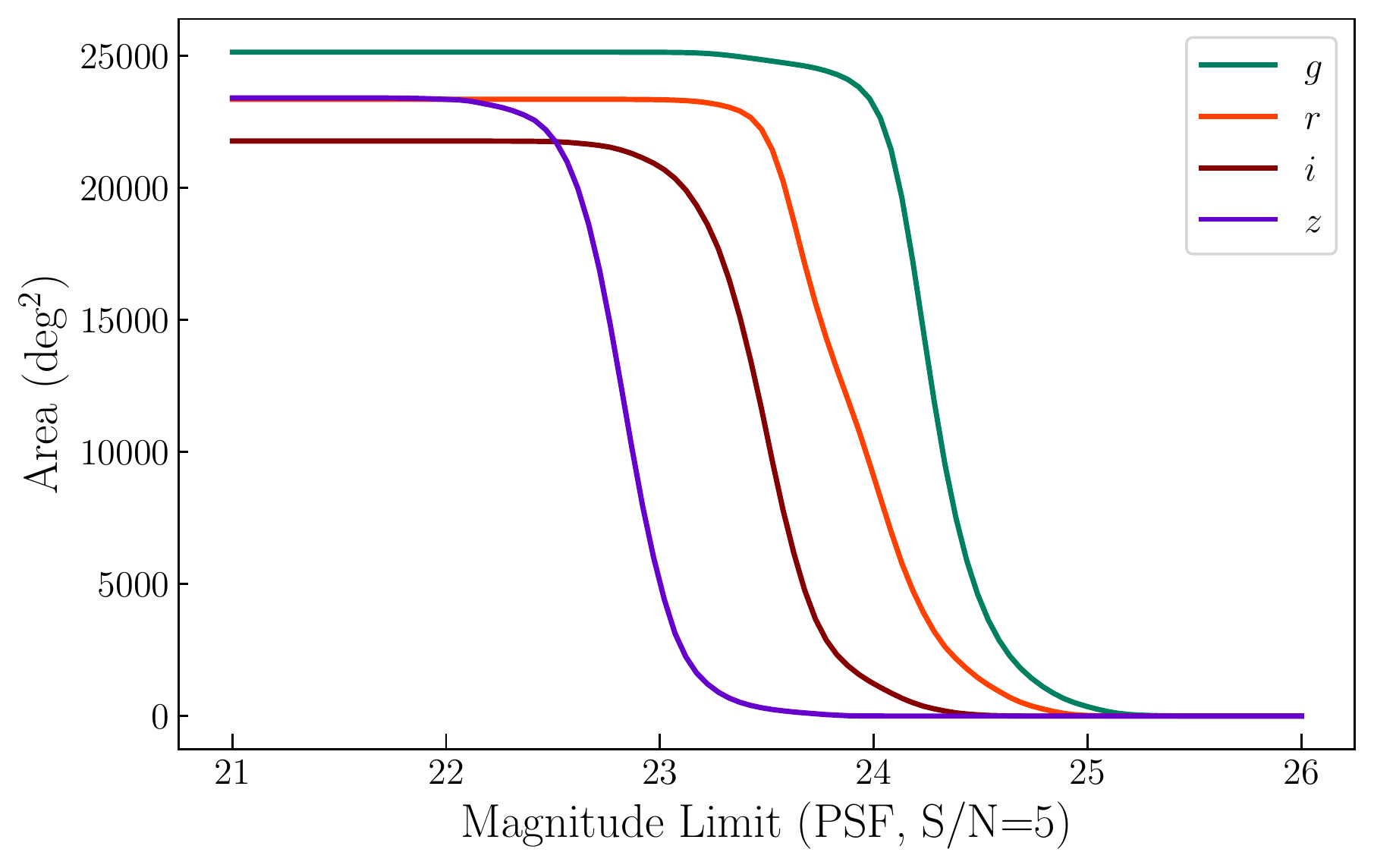}
    \caption{(Left) Distribution of PSF magnitude limits for point-like sources at S/N=5. The double-peaked structure in $r$ band comes from the different exposure times used in DES and DECaLS. (Right) DELVE DR2 survey area in each band as a function of the limiting PSF magnitude (S/N=5). These distributions look similar when calculated from the \magauto limiting magnitude for all sources but are shifted brighter by $\roughly 0.4 \magn$.
    }
    \label{fig:maglim}
\end{figure*}

\subsection{Absolute Photometric Calibration}
\label{sec:photabs}

The photometry of DELVE DR2 is tied to the $AB$ magnitude system \citep{Oke:1983} via the HST CalSpec standard star C26202. 
Within the DES footprint, the DES FGCM zeropoints are directly tied to C26202 as described in Section 4.2.2 of \citet{DES-DR2:2021}.
Outside the DES footprint, the calibration is tied more indirectly to C26202 via the zeropoints of the ATLAS Refcat2 transformation equations, which were adjusted to match DES DR2 (see Appendix A of \citealt{Drlica-Wagner:2021}).
Due to this procedure, DELVE DR2 cannot have a better absolute calibration accuracy than DES DR2, which sets a lower limit on the statistical uncertainty of $2.2 \mmag$ per band and a systematic uncertainty of $11$ to $12 \mmag$ per band (see Table 1 of \citealt{DES-DR2:2021}).
The global offset seen between the PS1 and SkyMapper regions of ATLAS Refcat2 when compared to \Gaia EDR3 suggests that the absolute calibration cannot be better than $10 \mmag$.
Combining the maximum systematic uncertainty on the absolute calibration from DES DR2 and the DELVE DR2 offset relative to \Gaia EDR3, we estimate that the absolute photometric accuracy of DELVE DR2 is $\lesssim \photabs \mmag$.

DELVE performed dedicated observations of the CalSpec standard star SDSS151421 during twilight hours in 2020. 
These observations were not used to set the absolute calibration of DELVE DR2, and they can instead be used to validate our estimate of the absolute calibration uncertainty.
We find that the median offsets between the DELVE PSF magnitudes and the CalSpec STIS magnitudes for SDSS151421 are $\Delta{g}{=}\photabsg$, $\Delta{r}{=}\photabsr$, $\Delta{i}{=}\photabsi$, and $\Delta{z}{=}\photabsz$\,mmag with a scatter of $\roughly 6$\,mmag.
Similar analyses performed by DES found $\roughly 10\mmag$ offsets when comparing the DES photometry to several CalSpec standard stars and DA white dwarfs within the DES footprint \citep{DES-DR2:2021}.
Based on these comparisons, we maintain the stated absolute calibration accuracy of $\lesssim \photabs \mmag$.

\subsection{Photometric Depth}
\label{sec:depth}

\begin{deluxetable}{l c c c c c}
\tablewidth{0pt}
\tabletypesize{\footnotesize}
\tablecaption{ DELVE DR2 median depth estimates. } 
\label{tab:depth}
\tablehead{
\colhead{Measurement} & &\multicolumn{4}{c}{Magnitude Limit}  \\[-0.5em]
                      & & $g$ & $r$ & $i$ & $z$   \\[-0.25em]
                      & & (mag) & (mag) & (mag) & (mag)
}
\startdata
\var{MAG\_PSF~} (S/N=5)  & & \maglimpsfg & \maglimpsfr & \maglimpsfi & \maglimpsfz  \\
\var{MAG\_PSF~} (S/N=10) & & \maglimpsfteng & \maglimpsftenr & \maglimpsfteni & \maglimpsftenz  \\
\var{MAG\_AUTO} (S/N=5)  & & \maglimautog & \maglimautor & \maglimautoi & \maglimautoz  \\
\var{MAG\_AUTO} (S/N=10) & & \maglimautoteng & \maglimautotenr & \maglimautoteni & \maglimautotenz \\
\enddata
\tablecomments{The \var{MAG\_PSF} depth is estimated from  point-like sources, while the \var{MAG\_AUTO} depth is estimated from all DELVE DR2 sources. Both \var{MAG\_PSF} and \var{MAG\_AUTO} are estimated from the best exposure of each object (see \secref{depth}).}
\end{deluxetable}

The photometric depth of DELVE DR2 can be assessed in several ways.
One common metric is to determine the magnitude at which a fixed signal-to-noise ratio (S/N) is achieved \citep[e.g.,][]{Rykoff:2015}.
The statistical magnitude uncertainty is related to the S/N calculated from the flux, $F/\delta F$, via propagation of uncertainties and Pogson's law \citep{Pogson:1856},
\begin{equation}
\delta m = \frac{2.5}{\ln 10} \frac{\delta F}{F}.
\end{equation}
\noindent Using this equation, we estimate the magnitude at which DELVE DR2 achieves S/N=5 ($\delta m \approx 0.2171$) and S/N=10 ($\delta m \approx 0.1085$).
We calculate these magnitude limits for point-like sources using \magpsf and for all sources using \magauto.
For each magnitude and S/N combination, we select objects and interpolate the relationship between $m$ and ${\rm median}(\delta m)$ in $\roughly 12 \amin^2$ \healpix pixels ($\nside = 1024$).
The resulting median magnitude limits estimated over the DELVE DR2 footprint are shown in \tabref{depth}.
We show histograms of the \magpsf magnitude limit for point-like sources at S/N=5 in the left panel of \figref{maglim}.
In the right panel of \figref{maglim} we show the DELVE DR2 area as a function of depth in each band.
The magnitude limits as a function of location on the sky are shown in \appref{depth}. 
Due to the catalog-level coaddition process, the depth of DELVE DR2 is set by the single best exposure in any region of the sky. 
This means that the depth of DELVE DR2 is very similar to that of DELVE DR1 \citep{Drlica-Wagner:2021} and significantly shallower than DES DR2 even in the overlapping DES region \citep{DES-DR2:2021}.
At bright magnitudes, the DECam CCDs will saturate at $g = 15.2$, $r = 15.7$, $i = 15.8$, and $z = 15.5$ for point sources observed in a $90\second$ exposure with median seeing \citep{DES-DR2:2021}.
While $\roughly 85\%$ of the exposures included in DELVE DR2 have exposure times of $\lesssim 90\second$, there are some regions with longer exposure times where saturation will occur at fainter magnitudes.
Therefore, objects detected by \sextractor with the saturation flag bit set were removed from the DELVE DR2 catalog production.


\subsection{Object Classification}
\label{sec:classification}

\begin{figure*}[t!]
    \centering
    \includegraphics[width=0.49\textwidth]{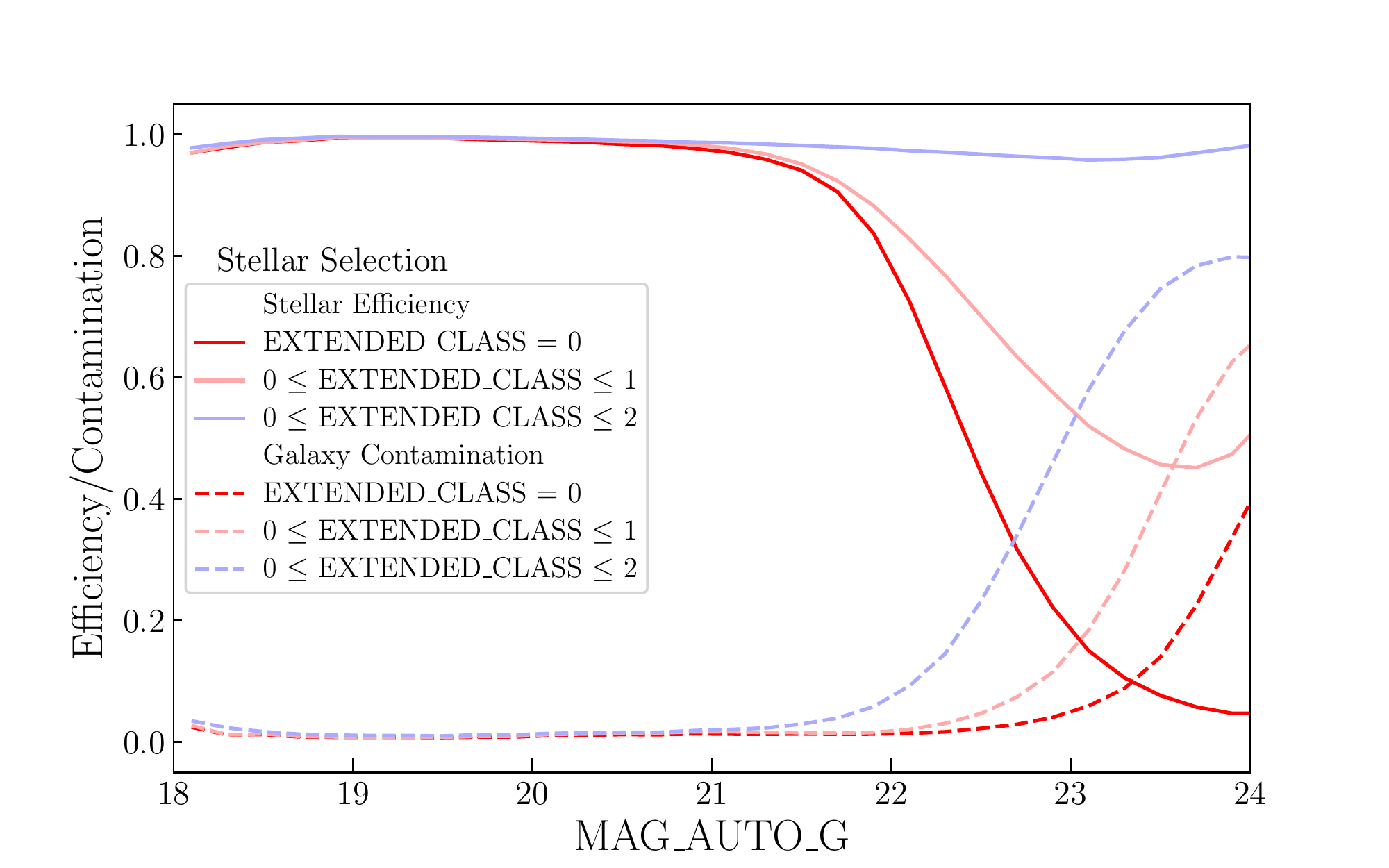}
    \includegraphics[width=0.49\textwidth]{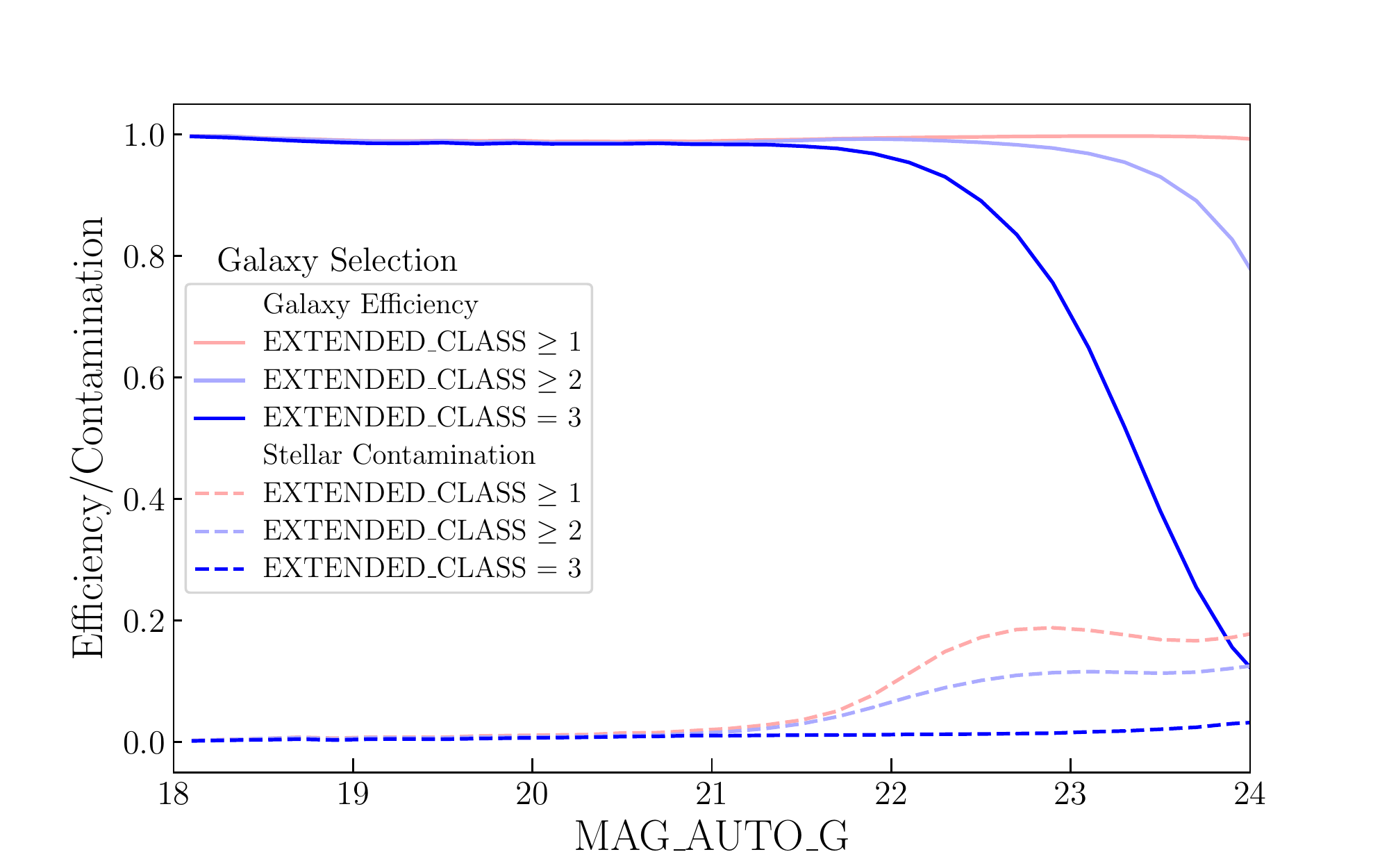}
    \caption{
      DELVE DR2 star/galaxy classification performance as a function of magnitude estimated from matched objects in the wide layer of HSC-SSP PDR3. 
      \emph{Left:} Stellar efficiency and galaxy contamination for several stellar samples based on \extclass[g].
      \emph{Right:} Galaxy efficiency and stellar contamination as a function of magnitude for several galaxy samples based on \extclass[g].
      \label{fig:stargal}
}
\end{figure*}

DELVE DR2 includes the \SExtractor \spreadmodel parameter, which can be used to separate spatially extended galaxies from point-like stars and quasars \citep[\eg,][]{Desai:2012}.
Following DES \citep[e.g.,][]{DES-DR1:2018,DES-DR2:2021} and DELVE DR1 \citep{Drlica-Wagner:2021}, we define \extclass parameters as a sum of several Boolean conditions,
{\par\footnotesize\begin{align}\begin{split}
\var{ext}&\var{ended\_class\_g} = \\
~&((\spreadmodel[g] + 3\, \spreaderrmodel[g]) > 0.005) \\
+&((\spreadmodel[g] +     \spreaderrmodel[g]) > 0.003) \\
+&((\spreadmodel[g] -     \spreaderrmodel[g]) > 0.003).
\end{split}\end{align}}

\noindent When true, each Boolean condition adds one unit to the classifier such that an \extclass value of 0 indicates high-confidence stars, 1 is likely stars, 2 is likely galaxies, and 3 is high-confidence galaxies.
Objects that lack coverage in a specific band or where the \spreadmodel fit failed are set to a sentinel value of $-9$.
We calculate \extclass values similarly for each band; however, we recommend the use of the $g$-band classifier, \extclass[G], since the $g$ band has the widest coverage and deepest limiting magnitude. 

In \figref{stargal}, we characterize the performance of \extclass[g] as a function of magnitude by matching DELVE DR2 objects to data from the W04 (WIDE12H+GAMA15H) equatorial field of the wide layer of HSC-SSP PDR3 \citep{HSC-PDR3}. 
To improve uniformity, we select only overlapping regions where the S/N = 5 limiting PSF magnitude from DELVE is representative of the DELVE DR2 survey (magnitude limit of $24 < g < 24.5$; \appref{depth}).
The superior image quality ($i$-band PSF FWHM $\roughly 0\farcs61$) and depth ($i \sim 26.2 \magn$) of the wide layer of HSC-SSP PDR3 enable robust tests of star--galaxy separation in DELVE DR2.
The matched data set covers $\roughly 394 \deg^2$ and contains $\roughly 9.6$ million matched objects.
Following previous analyses \citep{DES-DR1:2018,Drlica-Wagner:2021}, we select point-like sources from HSC-SSP PDR3 based on the difference between the $i$-band PSF and model magnitudes of sources,
{\par\footnotesize\begin{align}\begin{split}
\var{hsc}&\var{\_stars} = \\
& ((\var{i\_psfflux\_mag} - \var{i\_cmodel\_mag}) < 0.03) \\
& ||~ (~ ( (\var{i\_psfflux\_mag} - \var{i\_cmodel\_mag}) < 0.1) \\
& ~~~ \& ~ (\var{i\_psfflux\_mag < 22})~).
\end{split}\end{align}}
\noindent This scheme requires that the PSF and model magnitudes are very similar for fainter sources, while the agreement is relaxed for brighter sources.
This selection results in $\roughly 7.1$ million matched objects classified as galaxies and $\roughly 2.5$ million matched objects classified as stars.
We use these objects to evaluate the differential performance of DELVE DR2 \extclass[g] as a function of magnitude in \figref{stargal}.
A nominal stellar sample ($0 \leq \extclass[g] \leq 1$) contains $\roughly \approxnstars$ objects, while a nominal galaxy sample ($2 \leq \extclass[g]$) contains $\roughly \approxngals$ objects.
We report the integrated efficiency and contamination of these samples over the magnitude range $19 \leq \magauto[g] \leq 22$\,mag in \tabref{summary}.

\begin{figure*}[t!]
    \centering
    \includegraphics[width=0.49\textwidth]{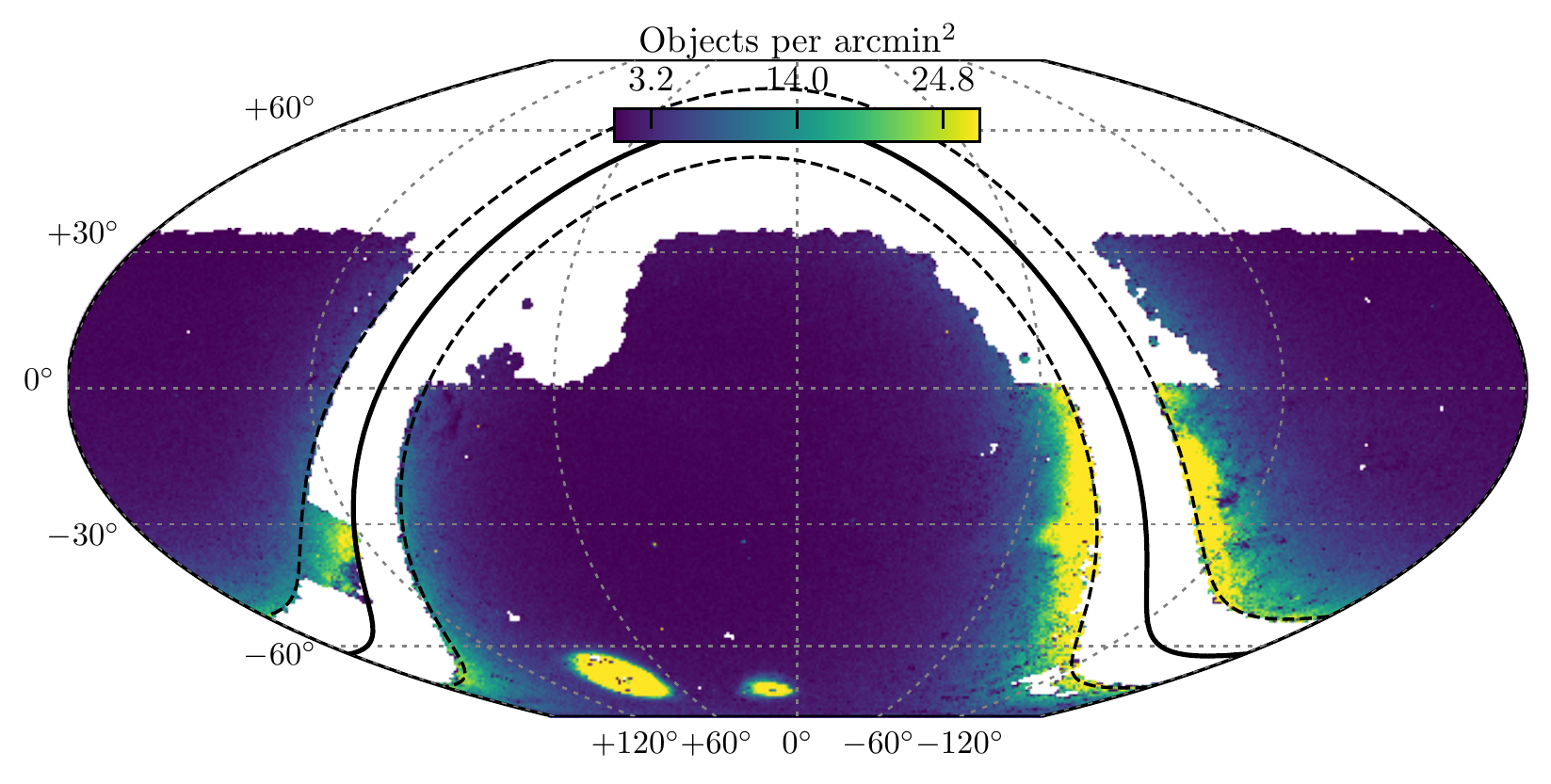}
    \includegraphics[width=0.49\textwidth]{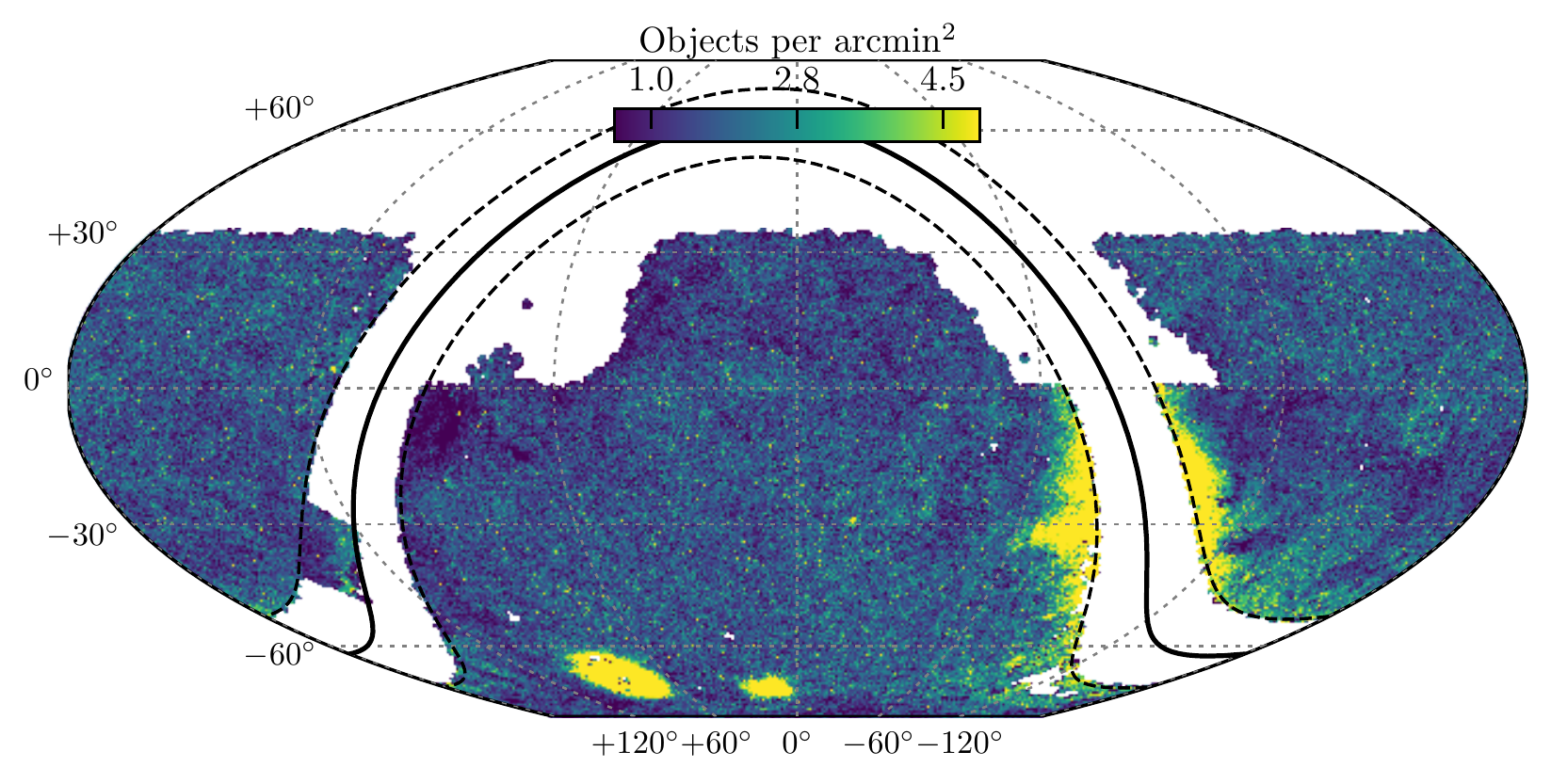}
    \caption{\label{fig:stargalmap} \textit{Left}: Stellar density map created with the $\extclass[g] = 0$ (high-confidence stars) selection described in \secref{classification}. \textit{Right}: Analogous galaxy counts map created with the $\extclass[G] = 3$ (high-confidence galaxies) selection. The region of lower galaxy density toward the northeast of the footprint can be attributed to higher interstellar extinction, which is not corrected for in this map. Color range units are number of objects per arcmin$^2$. Both maps apply a magnitude threshold of $\magauto[g] < 22$.}
\end{figure*}

The spatial number density of high-confidence stars (\extclass[g] = 0) and high-confidence galaxies (\extclass[g] = 3) are shown in \figref{stargalmap}.
The stellar density map clearly shows increasing stellar density toward the Galactic plane, as well as the high stellar density associated with the LMC and SMC. 
The galaxy density map is dominated by the large-scale clustering of galaxies at high Galactic latitudes, but stellar contamination is apparent close to the Galactic bulge, LMC, and SMC.
These maps have had a magnitude cut applied at $\var{mag\_auto\_i < 22}$ and have not been corrected for interstellar extinction, so some apparent variations in depth come from the extinction while others come from actual variations in depth over the footprint.

\subsection{Known Issues}
\label{sec:issues}

\begin{enumerate}
\item The DESDM pipeline was designed for galaxy photometry at high Galactic latitudes. Sky subtraction and deblending suffer in regions of high stellar density. This leads to degraded photometry and object classification in these regions, most notably close to the Galactic plane and the Magellanic Clouds (\figref{stargalmap}).
\item The star-galaxy classification efficiency varies over the footprint in a way that is found to correlate with imaging depth and object density. Care should be taken in regions of high density and/or spatially variable depth.
\item While the impact of scattered light from bright stars and failures in the sky background estimation have been mitigated in DELVE DR2 (\secref{data}), some localized, low-level catalog contamination does remain. The effects of scattered light may be further mitigated through the use of more advanced identification algorithms \citep[e.g.,][]{Tanoglidis:2022}.
\item Spatial coverage maps were created at a resolution of $\nside = 16384$, corresponding to linear pixel dimensions of $\roughly 13\arcsec$. Thus, there are a small number of catalog objects that reside outside the coverage maps due to the slight inaccuracy at the CCD boundaries. These objects reside at the edges of the DELVE footprint and are $<0.0001\%$ of the catalog.
\end{enumerate}


\section{Data Access}
\label{sec:access}

Access to DELVE DR2 is provided through the Astro Data Lab \citep{Fitzpatrick:2016,Nikutta:2020},\footnote{\url{https://datalab.noirlab.edu}} part of the Community Science and Data Center (CSDC) hosted by NOIRLab.
DELVE DR2 includes a main object table consisting of photometric measurements for $\roughly \approxnobjs$ objects.
In addition, the Astro Data Lab has computed cross-match tables between the DELVE DR2 catalog and catalogs from AllWISE, \Gaia EDR3, NSC DR2, SDSS DR16, and unWISE DR1  \citep[][]{Cutri:2021,Gaia:2021,Nidever:2021a,Ahumada:2020,Schlafly:2019}. 
These cross-match tables and their reverse counterparts are served alongside the DELVE DR2 main object table at the Astro Data Lab (see \appref{tables}).
The DELVE DR2 catalog data can be accessed via both a Table Access Protocol (TAP)\footnote{\url{http://ivoa.net/documents/TAP}} service and from direct PostgreSQL queries via web-based, command-line, and programmatic query interfaces.
In addition, the Astro Data Lab provides an image cutout service, built on the Simple Image Access (SIA) protocol, that can be used to access versions of the DELVE DR2 imaging data processed with the DECam Community Pipeline \citep{Valdes:2014}. 
More detailed information on accessing the DELVE DR2 data can be found on the Astro Data Lab website.\footnote{\url{https://datalab.noirlab.edu/delve}}


\section{Summary}
\label{sec:summary}

DELVE seeks to study the physics of dark matter and galaxy formation by observing resolved dwarf galaxies and stellar substructures in the Local Volume.
To do so, DELVE has set out to complete contiguous deep imaging coverage of the southern high Galactic latitude sky.
DELVE DR2 combines new observations with archival DECam data to cover $>20{,}000 \deg^2$ individually in $g$, $r$, $i$, $z$ and $\roughly \approxarea \deg^2$ in all four bands simultaneously.
The DELVE DR2 catalog contains PSF and automatic aperture measurements for $\roughly \approxnobjs$ astronomical objects with a $5\sigma$ PSF depth of $g=\maglimpsfg, r=\maglimpsfr, i=\maglimpsfi, z=\maglimpsfz \magn$ (\tabref{summary}).
The DELVE DR2 data products are accessible through the NOIRLab Astro Data Lab.

As of 2022 January, DELVE has completed $\roughly 80\%$ of its 126 nights of scheduled DECam observing. 
Additional DECam observations will increase the coverage, uniformity, and depth of future DELVE catalogs.
Furthermore, we expect that future DELVE data releases will include products derived from image coaddition, as well as deeper targeted regions of the DELVE footprint.
We anticipate that DELVE DR2 and future DELVE data releases will be a valuable resource for the community in advance of the Vera C.\ Rubin Observatory Legacy Survey of Space and Time.


\section{Acknowledgments}
The DELVE project is partially supported by Fermilab LDRD project L2019-011, the NASA Fermi Guest Investigator Program Cycle 9 grant 91201, and the National Science Foundation (NSF) under grant AST-2108168.
This work is supported by the Fermilab Visiting Scholars Award Program from the Universities Research Association.

ABP acknowledges support from NSF grant AST-1813881.
JLC acknowledges support from NSF grant AST-1816196.
JDS acknowledges support from NSF grant AST-1714873.
SRM acknowledges support from NSF grant AST-1909497.
DJS acknowledges support from NSF grants AST-1821967 and AST-1813708.
DC acknowledges support from NSF grant AST-1814208.
BMP acknowledges support from the NSF Astronomy and Astrophysics Postdoctoral Fellowship under award AST-2001663.
SM acknowledges support from the NSF Graduate Research Fellowship under grant DGE-1656518.
DMD acknowledges financial support from the State Agency for Research
of the Spanish MCIU through the ``Centre of Excellence Severo Ochoa''
award for the Instituto de Astrofísica de Andaluc\'ia (SEV-2017-0709).
CPMB and MRLC acknowledge support from the European Research Council (ERC) under the European Union's Horizon 2020 research and innovation programme (grant agreement no.\ 682115).
LSS acknowledges the financial support from FAPESP through the grant \#2020/03301-5.
JACB acknowledges support from ANID FONDECYT Regular 1220083.

Funding for the DES Projects has been provided by the U.S. Department of Energy, the U.S. National Science Foundation, the Ministry of Science and Education of Spain, 
the Science and Technology Facilities Council of the United Kingdom, the Higher Education Funding Council for England, the National Center for Supercomputing 
Applications at the University of Illinois at Urbana-Champaign, the Kavli Institute of Cosmological Physics at the University of Chicago, 
the Center for Cosmology and Astro-Particle Physics at the Ohio State University,
the Mitchell Institute for Fundamental Physics and Astronomy at Texas A\&M University, Financiadora de Estudos e Projetos, 
Funda{\c c}{\~a}o Carlos Chagas Filho de Amparo {\`a} Pesquisa do Estado do Rio de Janeiro, Conselho Nacional de Desenvolvimento Cient{\'i}fico e Tecnol{\'o}gico and 
the Minist{\'e}rio da Ci{\^e}ncia, Tecnologia e Inova{\c c}{\~a}o, the Deutsche Forschungsgemeinschaft and the Collaborating Institutions in the Dark Energy Survey. 

The Collaborating Institutions are Argonne National Laboratory, the University of California at Santa Cruz, the University of Cambridge, Centro de Investigaciones Energ{\'e}ticas, 
Medioambientales y Tecnol{\'o}gicas-Madrid, the University of Chicago, University College London, the DES-Brazil Consortium, the University of Edinburgh, 
the Eidgen{\"o}ssische Technische Hochschule (ETH) Z{\"u}rich, 
Fermi National Accelerator Laboratory, the University of Illinois at Urbana-Champaign, the Institut de Ci{\`e}ncies de l'Espai (IEEC/CSIC), 
the Institut de F{\'i}sica d'Altes Energies, Lawrence Berkeley National Laboratory, the Ludwig-Maximilians Universit{\"a}t M{\"u}nchen and the associated Excellence Cluster Universe, 
the University of Michigan, NSF's NOIRLab, the University of Nottingham, The Ohio State University, the University of Pennsylvania, the University of Portsmouth, 
SLAC National Accelerator Laboratory, Stanford University, the University of Sussex, Texas A\&M University, and the OzDES Membership Consortium.

The DES data management system is supported by the National Science Foundation under Grant Numbers AST-1138766 and AST-1536171.
The DES participants from Spanish institutions are partially supported by MICINN under grants ESP2017-89838, PGC2018-094773, PGC2018-102021, SEV-2016-0588, SEV-2016-0597, and MDM-2015-0509, some of which include ERDF funds from the European Union. IFAE is partially funded by the CERCA program of the Generalitat de Catalunya.
Research leading to these results has received funding from the European Research
Council under the European Union's Seventh Framework Program (FP7/2007-2013) including ERC grant agreements 240672, 291329, and 306478.
We  acknowledge support from the Brazilian Instituto Nacional de Ci\^encia
e Tecnologia (INCT) do e-Universo (CNPq grant 465376/2014-2).

Based in part on observations at Cerro Tololo Inter-American Observatory at NSF's NOIRLab, which is managed by the Association of Universities for Research in Astronomy (AURA) under a cooperative agreement with the National Science Foundation.

This work has made use of data from the European Space Agency (ESA) mission {\it Gaia} (\url{https://www.cosmos.esa.int/gaia}), processed by the {\it Gaia} Data Processing and Analysis Consortium (DPAC, \url{https://www.cosmos.esa.int/web/gaia/dpac/consortium}).
Funding for the DPAC has been provided by national institutions, in particular the institutions participating in the {\it Gaia} Multilateral Agreement.

This paper is based on data collected at the Subaru Telescope and retrieved from the HSC data archive system, which is operated by the Subaru Telescope and Astronomy Data Center (ADC) at NAOJ. Data analysis was in part carried out with the cooperation of Center for Computational Astrophysics (CfCA), NAOJ. We are honored and grateful for the opportunity of observing the Universe from Maunakea, which has the cultural, historical and natural significance in Hawaii. 

This manuscript has been authored by Fermi Research Alliance, LLC under Contract No. DE-AC02-07CH11359 with the U.S. Department of Energy, Office of Science, Office of High Energy Physics.


\facilities{Blanco (DECam), Astro Data Lab, \Gaia, Subaru (HSC)} 

\software{
\code{astropy} \citep{astropy:2018},
\code{fitsio},\footnote{\url{https://github.com/esheldon/fitsio}}
\healpix \citep{Gorski:2005},\footnote{\url{http://healpix.sourceforge.net}}
\code{healpy} \citep{Zonca:2019},\footnote{\url{https://github.com/healpy/healpy}}
\code{healsparse},\footnote{\url{https://healsparse.readthedocs.io/en/latest/}}
\code{matplotlib} \citep{Hunter:2007},
\code{numpy} \citep{NumPy:2020},
\PSFEx \citep{Bertin:2011},
\code{scipy} \citep{Scipy:2020},
\scamp \citep{Bertin:2006}, 
\code{skymap},\footnote{\url{https://github.com/kadrlica/skymap}}
\SExtractor \citep{Bertin:1996}
}


\appendix
\clearpage

\section{DECam Data}
\label{app:propid}

DELVE DR2 combines DECam observations acquired by 278 programs. 
These programs and the number of exposures they each contributed to DELVE DR2 are listed in \tabref{propid}.

\startlongtable{
\begin{deluxetable}{| c c c | c c c | c c c |}
\tablecolumns{3}
\tabletypesize{\scriptsize}
\tablecaption{\label{tab:propid}
DECam data included in DELVE DR2}
\tablehead{
\colhead{Prop.ID}  & \colhead{PI} & \colhead{$N_{\rm exp}$} &\colhead{Prop.ID}  & \colhead{PI} & \colhead{$N_{\rm exp}$} & \colhead{Prop.ID}  & \colhead{PI} & \colhead{$N_{\rm exp}$}}
\startdata
2012B-0001 & Josh Frieman & 63656 & 2018A-0909 & Thomas H Puzia & 121 & 2012B-0620 & Jeremy Mould & 23\\
2014B-0404 & David Schlegel & 28823 & 2015A-0631 & Alfredo Zenteno & 120 & 2021A-0010 & Travis Rector & 23\\
2019A-0305 & Alex Drlica-Wagner & 12459 & 2017B-0312 & Bryan Miller & 119 & 2019B-0080 & Casey Papovich & 23\\
2018A-0386 & Alfredo Zenteno & 3029 & 2019A-0265 & Douglas P Finkbeiner & 119 & 2013A-0737 & Scott Sheppard & 22\\
2013B-0440 & David Nidever & 2753 & 2016B-0124 & Edo Berger & 111 & 2016A-0622 & Paulo Lopes & 22\\
2019A-0272 & Alfredo Zenteno & 2452 & 2013B-0421 & Armin Rest & 107 & 2016A-0191 & Armin Rest & 22\\
2017A-0260 & Marcelle Soares-Santos & 2297 & 2020A-0058 & Kathy Vivas & 107 & 2012B-3001 & Emmanuel Bertin & 21\\
2021A-0149 & Alfredo Zenteno & 1886 & 2015B-0606 & Katharine Lutz & 106 & 2015A-0322 & R Michael Rich & 21\\
2016A-0366 & Keith Bechtol & 1870 & 2020A-0402 & ------ & 102 & 2019A-0240 & ------ & 20\\
2019B-0323 & Alfredo Zenteno & 1586 & 2015B-0187 & Edo Berger & 98 & 2018A-0371 & Sangeeta Malhotra & 20\\
2017A-0388 & Alfredo Zenteno & 1432 & 2017B-0906 & Dougal Mackey & 97 & 2012B-0625 & Sarah Sweet & 20\\
2018A-0242 & Keith Bechtol & 1423 & 2017A-0298 & Brad Tucker & 96 & 2014A-0496 & Aren Heinze & 20\\
2020A-0399 & Alfredo Zenteno & 1387 & 2018A-0159 & Kathy Vivas & 96 & 2019B-0256 & Michael M Shara & 20\\
2021A-0275 & Armin Rest & 1336 & 2014A-0339 & Jonathan Hargis & 95 & 2012B-0621 & Loren Bruns & 19\\
2018A-0273 & William Dawson & 1192 & 2014A-0622 & Iraklis Konstantopoulos & 92 & 2014B-0265 & Ian Dell'Antonio & 19\\
2018A-0913 & Brad Tucker & 1086 & 2020A-0910 & Thomas H Puzia & 91 & 2015B-0175 & Anton Koekemoer & 19\\
2013A-0741 & David Schlegel & 997 & 2018A-0380 & Armin Rest & 90 & 2013B-0627 & Gastao B Lima Neto & 18\\
2019A-0308 & Ian Dell'Antonio & 944 & 2019B-0403 & Clara Martinez-Vazquez & 87 & 2014A-0621 & Dougal Mackey & 18\\
2013A-0327 & Armin Rest & 900 & 2014A-0239 & Mark Sullivan & 83 & 2013A-9999 & Alistair Walker & 18\\
2014A-0624 & Helmut Jerjen & 817 & 2018B-0941 & Alistair Walker & 82 & 2014A-0634 & David James & 17\\
2017B-0279 & Armin Rest & 790 & 2018A-0137 & Jeffrey Cooke & 76 & 2014B-0611 & Douglas P Geisler & 17\\
2013A-0214 & Maureen Van Den Berg & 772 & 2014A-0429 & Douglas P Finkbeiner & 74 & 2013A-0386 & Paul Thorman & 16\\
2013A-0360 & Anja von der Linden & 737 & 2017B-0239 & Keith Bechtol & 72 & 2014A-0073 & Mukremin Kilic & 16\\
2013A-0724 & Lori Allen & 708 & 2013B-0612 & Julio Chaname & 71 & 2015A-0618 & Chris Lidman & 15\\
2018A-0914 & Martin Makler & 704 & 2019A-0065 & Yue Shen & 70 & 2014B-0375 & Armin Rest & 15\\
2015A-0608 & Francisco Forster & 638 & 2018B-0340 & Herve Bouy & 70 & 2014A-0386 & Ian Dell'Antonio & 15\\
2014A-0415 & Anja von der Linden & 604 & 2015A-0151 & Annalisa Calamida & 70 & 2014B-0610 & Julio Chaname & 14\\
2014A-0306 & Xinyu Dai & 559 & 2014A-0348 & Haojing Yan & 68 & 2012B-3005 & Knut Olsen & 14\\
2015A-0616 & Helmut Jerjen & 467 & 2017B-0285 & Armin Rest & 68 & 2019A-0337 & David E Trilling & 14\\
2016B-0909 & Camila Navarrete & 462 & 2017B-0078 & Herve Bouy & 68 & 2014B-0064 & Mukremin Kilic & 14\\
2013A-0614 & Sarah Sweet & 460 & 2019A-0235 & ------ & 67 & 2016A-0337 & Genaro Suarez Castro & 12\\
2016B-0301 & Armin Rest & 439 & 2018B-0905 & Stree Oh & 66 & 2017A-0951 & Kathy Vivas & 12\\
2019B-1014 & Felipe Olivares & 437 & 2014A-0632 & Tiago Gon\c{c}alves & 65 & 2013A-0351 & Arjun Dey & 12\\
2015A-0620 & Ana Bonaca & 430 & 2020A-0353 & Eric Peng & 65 & 2013B-0615 & Julio Carballo-Bello & 12\\
2014A-0035 & Herve Bouy & 427 & 2016A-0384 & Jacqueline McCleary & 64 & 2015A-0062 & Linda French & 12\\
2018B-0271 & Douglas P Finkbeiner & 424 & 2014A-0480 & R Michael Rich & 63 & 2019B-1013 & Thomas H Puzia & 12\\
2019A-0910 & Dougal Mackey & 424 & 2014A-0313 & Kathy Vivas & 62 & 2015A-0610 & Cesar Fuentes & 12\\
2015A-0110 & Thomas De Boer & 379 & 2015B-0307 & Armin Rest & 61 & 2014B-0613 & Jeffrey Cooke & 11\\
2014A-0270 & Carl J Grillmair & 363 & 2018A-0206 & Abhijit Saha & 61 & 2014B-0614 & Iraklis Konstantopoulos & 11\\
2016A-0189 & Armin Rest & 359 & 2015A-0617 & David M Nataf & 60 & 2012B-0623 & Dougal Mackey & 10\\
2013A-0411 & David Nidever & 358 & 2017A-0210 & Alistair Walker & 60 & 2016A-0095 & Jeffrey Cooke & 10\\
2016A-0618 & Dougal Mackey & 349 & 2013B-0617 & Dougal Mackey & 59 & 2016A-0951 & ------ & 10\\
2020A-0908 & Felipe Olivares & 339 & 2013A-0529 & R Michael Rich & 59 & 2015A-0175 & Taran Esplin & 9\\
2014A-0608 & Francisco Forster & 335 & 2014B-0193 & Frederick M Walter & 58 & 2013B-0453 & Scott Sheppard & 9\\
2016A-0190 & Arjun Dey & 333 & 2017B-0103 & Wayne Barkhouse & 58 & 2018B-0327 & Sangeeta Malhotra & 9\\
2021A-0922 & Jose L Nilo Castellon & 332 & 2019B-0042 & Herve Bouy & 57 & 2015A-0609 & Julio Carballo-Bello & 9\\
2020B-0241 & Alfredo Zenteno & 330 & 2014A-0613 & David Rodriguez & 57 & 2019A-0911 & Jeffrey Cooke & 9\\
2018A-0251 & Douglas P Finkbeiner & 324 & 2019A-0101 & Patrick M Hartigan & 57 & 2020B-0053 & Dillon Brout & 8\\
2018A-0276 & Ian Dell'Antonio & 304 & 2016A-0614 & Thomas H Puzia & 57 & 2017B-0330 & Sangeeta Malhotra & 7\\
2014A-0412 & Armin Rest & 303 & 2019B-0910 & Yue Shen & 55 & 2013A-0455 & Scott Sheppard & 7\\
2013A-0719 & Abhijit Saha & 291 & 2017B-0163 & Prashin Jethwa & 54 & 2012B-0416 & David Nidever & 7\\
2019A-0205 & Daniel Goldstein & 290 & 2013A-0612 & Yun-Kyeong Sheen & 53 & 2017B-0199 & Anton Koekemoer & 7\\
2018A-0215 & Jeffrey Carlin & 289 & 2017A-0913 & Luidhy Santana da Silva & 51 & 2013A-0609 & Douglas P Geisler & 7\\
2014A-0620 & Andrew Casey & 287 & 2014A-0610 & Matthew Taylor & 50 & 2020B-0021 & Haojing Yan & 7\\
2015A-0306 & Eduardo Balbinot & 280 & 2015A-0371 & Armin Rest & 50 & 2017A-0366 & Sangeeta Malhotra & 7\\
2014B-0244 & Anja von der Linden & 280 & 2016B-0173 & Anton Koekemoer & 49 & 2017B-0253 & Jeffrey Carlin & 6\\
2019B-0371 & Marcelle Soares-Santos & 280 & 2017A-0909 & Jeffrey Cooke & 49 & 2014A-0399 & Christopher Johnson & 6\\
2016B-0905 & Helmut Jerjen & 276 & 2015A-0615 & Brendan McMonigal & 49 & 2015B-0314 & Brad Tucker & 5\\
2017A-0914 & Grant Tremblay & 274 & 2017A-0308 & Annalisa Calamida & 48 & 2020A-0415 & Armin Rest & 5\\
2016A-0397 & Anja von der Linden & 263 & 2017A-0389 & Armin Rest & 48 & 2014A-0640 & Amy Mainzer & 5\\
2017A-0060 & Denija Crnojevic & 261 & 2014B-0609 & Roberto R Munoz & 47 & 2014B-0071 & Sarah Sonnett & 5\\
2017A-0281 & Monika D Soraisam & 256 & 2018A-0912 & Attila Popping & 45 & 2015B-0607 & Jeffrey Cooke & 5\\
2017A-0916 & Julio Carballo-Bello & 242 & 2021A-0246 & ------ & 44 & 2019B-1012 & Jeffrey Cooke & 5\\
2020A-0335 & Lifan Wang & 242 & 2020A-0238 & Clara Martinez-Vazquez & 43 & 2017B-0307 & Scott Sheppard & 4\\
2017B-0907 & Ricardo Munoz & 228 & 2019A-0325 & Clara Martinez-Vazquez & 43 & 2012B-0451 & Scott Sheppard & 4\\
2015A-0630 & Thomas H Puzia & 218 & 2020A-0142 & Tom Shanks & 42 & 2015A-0614 & Jeffrey Cooke & 4\\
2016A-0327 & Douglas P Finkbeiner & 216 & 2014B-0608 & Yara Jaffe & 41 & 2013B-0325 & Kathy Vivas & 4\\
2018B-0122 & Armin Rest & 213 & 2017A-0911 & Ana Chies Santos & 39 & 2020B-0288 & Alexie Leauthaud & 4\\
2012B-0569 & Lori Allen & 206 & 2020A-0909 & Patricia Arevalo & 39 & 2012B-0624 & Aaron Robotham & 4\\
2019A-0915 & Jose Pena & 191 & 2016A-0004 & Ana Bonaca & 38 & 2012B-3002 & Josh Bloom & 4\\
2015A-0619 & Thiago Goncalves & 186 & 2014A-0157 & Andrej Favia & 38 & 2015B-0603 & Leopoldo Infante & 4\\
2014A-0327 & Armin Rest & 183 & 2012B-0363 & Josh Bloom & 38 & 2015A-0177 & Cristian Eduard Rusu & 3\\
2018A-0059 & Herve Bouy & 182 & 2016A-0068 & Thomas Deboer & 38 & 2012B-0448 & Paul Thorman & 3\\
2015A-0163 & Carl J Grillmair & 179 & 2015B-0191 & Sarah Rice & 37 & 2014B-0378 & Armin Rest & 3\\
2018A-0911 & Francisco Forster & 174 & 2014A-0255 & Anton Koekemoer & 35 & 2013A-0613 & Ricardo Munoz & 3\\
2017B-0110 & Edo Berger & 174 & 2017B-0951 & Kathy Vivas & 35 & 2013A-0400 & Josh Bloom & 3\\
2016B-0910 & Thomas H Puzia & 174 & 2016B-0904 & Igor Andreoni & 33 & 2013A-0616 & Geraint Lewis & 2\\
2015A-0130 & Denija Crnojevic & 173 & 2019A-0060 & Herve Bouy & 33 & 2020A-0913 & Jeremy Mould & 2\\
2016B-0279 & Douglas P Finkbeiner & 170 & 2021A-0113 & Melissa L Graham & 33 & 2016A-0610 & Leopoldo Infante & 2\\
2013B-0614 & Ricardo Munoz & 167 & 2018A-0907 & Ricardo Munoz & 32 & 2013A-0608 & Ricardo Demarco & 2\\
2015A-0121 & Anja von der Linden & 160 & 2019B-1004 & Julio Chaname & 32 & 2014A-0191 & Hendrik Hildebrandt & 2\\
2019A-0917 & Paulo Lopes & 159 & 2012B-0506 & Daniel D Kelson & 32 & 2017B-0905 & Jeremy Mould & 2\\
2018A-0369 & Armin Rest & 156 & 2015A-0632 & Cesar Briceno & 31 & 2015B-0250 & Jonathan Hargis & 1\\
2017A-0918 & Alexandra Yip & 155 & 2013B-0531 & Eric Mamajek & 31 & 2012B-3016 & Scott Sheppard & 1\\
2013A-0611 & Dougal Mackey & 142 & 2018B-0904 & Lee Splitter & 30 & 2012B-0617 & Robert I Hynes & 1\\
2014B-0146 & Mark Sullivan & 141 & 2014A-0623 & Ken Freeman & 30 & 2013A-0610 & Mario Hamuy & 1\\
2014A-0256 & Kathleen Eckert & 136 & 2013A-0723 & Eric Mamajek & 28 & 2016A-0386 & Sangeeta Malhotra & 1\\
2015A-0205 & Eric Mamajek & 135 & 2019A-0315 & Matthew Penny & 28 & 2017A-0917 & Franz Bauer & 1\\
2014A-0321 & Marla Geha & 133 & 2013A-2101 & Alistair Walker & 28 & 2013B-0502 & Ian Dell'Antonio & 1\\
2017B-0904 & Paulo Lopes & 133 & 2015A-0107 & Claudia Belardi & 28 & 2013B-0613 & Roberto R Munoz & 1\\
2019A-0913 & Julio Carballo-Bello & 133 & 2013B-0438 & Casey Papovich & 26 & 2015A-0059 & Sarah Sonnett & 1\\
2015A-0397 & Armin Rest & 126 & 2013A-0621 & Matias Gomez & 25 & 2013A-0704 & Matt A Wood & 1\\
2019B-1010 & Jose Pena & 123 & 2016A-0104 & Mark Sullivan & 24 &  &  & \\
\enddata
\tablecomments{Programs are ordered by the number of exposures contributed. The largest single contributors to the DELVE DR2 data set are DES, DECaLS and the DELVE program itself. Programs with no principal investigator (PI) listed are generally Target-of-Opportunity (ToO) or multi-PI programs.}
\end{deluxetable}
}

\clearpage


\section{DELVE DR2 Tables}
\label{app:tables}
 
The DELVE DR2 catalog data are accessible through the \code{DELVE\_DR2.OBJECTS} table hosted by the Astro Data Lab.
This table includes the photometric properties assembled from a catalog-level co-add of the individual single-epoch measurements.
The table columns are described in \tabref{dr2_main}. 
In addition, cross-matches between objects in the DELVE DR2 catalog and objects within $1\farcs5$ from external catalogs are provided in individual tables:
\begin{itemize}[nosep]
\item \code{DELVE\_DR2.X1P5\_\_OBJECTS\_\_ALLWISE\_\_SOURCE} - AllWISE \citep{Cutri:2021}
\item \code{DELVE\_DR2.X1P5\_\_OBJECTS\_\_GAIA\_EDR3\_\_GAIA\_SOURCE} - \Gaia EDR3 \citep{Gaia:2021}
\item \code{DELVE\_DR2.X1P5\_\_OBJECTS\_\_NSC\_DR2\_\_OBJECT} - NSC DR2 \citep{Nidever:2021a}
\item \code{DELVE\_DR2.X1P5\_\_OBJECTS\_\_SDSS\_DR16\_\_SPECOBJ} - SDSS DR16 \citep{Ahumada:2020}
\item \code{DELVE\_DR2.X1P5\_\_OBJECTS\_\_UNWISE\_DR1\_\_OBJECT} - unWISE DR1 \citep{Schlafly:2019}
\end{itemize}
A template for the columns in these tables are described in \tabref{dr2_crossmatch}.
The schema for these tables are also described in detail on the Astro Data Lab website.

\begin{deluxetable}{l l c}
    	\centering
	\tablewidth{0pt}
	\tabletypesize{\scriptsize}
	\tablecaption{\code{DELVE.DR2\_MAIN} table description: \nobjs rows; 126 columns \label{tab:dr2_main}}
	\tablehead{
		\colhead{Column Name} & \colhead{Description} & \colhead{Columns}
	}
        \startdata
        QUICK\_OBJECT\_ID & Unique identifier for each object & 1 \\
	RA & Right ascension derived from the median position of each detection (deg) & 1 \\
	DEC & Declination derived from the median position of each detection (deg) & 1 \\
	GLON & Galactic longitude derived from RA,DEC (deg) & 1 \\
	GLAT & Galactic latitude derived from RA,DEC (deg) & 1 \\
	ELON & Ecliptic longitude derived from RA,DEC (deg) & 1 \\
	ELAT & Ecliptic latitude derived from RA,DEC (deg) & 1 \\
	A\_IMAGE\_\{G,R,I,Z\} & Semi-major axis of adaptive aperture in image coordinates (pix) & 4 \\
	B\_IMAGE\_\{G,R,I,Z\} & Semi-minor axis of adaptive aperture in image coordinates (pix) & 4 \\
	CCDNUM\_\{G,R,I,Z\} & CCD number for best exposure in each band & 4 \\
	CLASS\_STAR\_\{G,R,I,Z\} & Neural-network-based star--galaxy classifier (see \SExtractor manual for details) & 4 \\
	EBV & $E(B-V)$ value at the object location interpolated from the map of \citet{Schlegel:1998} & 1 \\
	EXPNUM\_\{G,R,I,Z\} & Exposure number for best exposure in each band & 4 \\
	EXPTIME\_\{G,R,I,Z\} & Shutter-open exposure time for best exposure in each band & 4 \\
	EXTENDED\_CLASS\_\{G,R,I,Z\} & Spread-model-based morphology class (see \secref{classification}) & 4\\
                                   & $-9$ unknown, 0 high-confidence star, 1 likely star, 2 likely galaxy, 3 high-confidence galaxy &  \\
	EXTINCTION\_\{G,R,I,Z\} & Interstellar extinction calculated from \citet{Schlegel:1998}. Subtract these columns from & 4 \\
                                & the magnitude columns to correct for extinction (see \secref{processing}). & \\
	FLAGS\_\{G,R,I,Z\} & \SExtractor flags for the best detection in each band & 4 \\
	HPX2048 & HEALPix index for each object in RING format at resolution $\nside=2048$ & 1 \\
	HTM9 & HTM Level-9 index & 1 \\
	MAG\_AUTO\_\{G,R,I,Z\} & Automatic aperture magnitude derived from the best exposure in each band  & 4 \\
	MAGERR\_AUTO\_\{G,R,I,Z\} & Automatic aperture magnitude uncertainty derived from the best exposure in each band & 4 \\
	MAG\_PSF\_\{G,R,I,Z\} &  PSF magnitude derived from the best exposure in each band  & 4 \\
	MAGERR\_PSF\_\{G,R,I,Z\} & PSF magnitude uncertainty derived from the best exposure in each band & 4 \\
	MJD\_OBS & Median Modified Julian Date of the observations that were used to determine the astrometric position & 1 \\
	NEPOCHS\_\{G,R,I,Z\} & Number of single-epoch detections for this object & 4 \\
        NEST4096 & \healpix index for each object in NEST format at resolution $\nside=4096$ & 1 \\
        RANDOM\_ID & Random ID in the range 0.0 to 100.0 for subsampling & 1 \\
        RING256 & \healpix index for each object in RING format at resolution $\nside=256$ & 1 \\
        SPREAD\_MODEL\_\{G,R,I,Z\} & Likelihood-based star--galaxy classifier \citep[][]{Desai:2012} & 4 \\
        SPREADERR\_MODEL\_\{G,R,I,Z\} & Likelihood-based star--galaxy classifier uncertainty \citep[][]{Desai:2012} & 4 \\
        T\_EFF\_\{G,R,I,Z\} & Effective exposure time scale factor for best exposure in each band \citep[][]{Neilsen:2016} & 4 \\
        THETA\_IMAGE\_\{G,R,I,Z\} & Position angle of automatic aperture in image coordinates (deg) & 4\\
        WAVG\_FLAGS\_\{G,R,I,Z\} & OR of \SExtractor flags from all detections in each band & 4 \\
        WAVG\_MAG\_AUTO\_\{G,R,I,Z\} & Weighted average of automatic aperture magnitude measurements in each band & 4 \\
        WAVG\_MAGERR\_AUTO\_\{G,R,I,Z\} & Sum in quadrature of the automatic aperture magnitude uncertainties in each band & 4 \\
        WAVG\_MAGRMS\_AUTO\_\{G,R,I,Z\} & Unbiased weighted standard deviation of the automatic aperture magnitude in each band & 4 \\
        WAVG\_MAG\_PSF\_\{G,R,I,Z\} & Weighted average of PSF magnitude measurements in each band & 4 \\
        WAVG\_MAGERR\_PSF\_\{G,R,I,Z\} & Sum in quadrature of the PSF magnitude uncertainties in each band & 4 \\
        WAVG\_MAGRMS\_PSF\_\{G,R,I,Z\} &  Unbiased weighted standard deviation of the PSF magnitude in each band & 4 \\
        WAVG\_SPREAD\_MODEL\_\{G,R,I,Z\} & Weighted average spread model in each band & 4 \\
        WAVG\_SPREADERR\_MODEL\_\{G,R,I,Z\} & Sum in quadrature of the spread model uncertainties in each band & 4 \\
        WAVG\_SPREADRMS\_MODEL\_\{G,R,I,Z\} & Unbiased weighted standard deviation of \var{spread\_model} in each band & 4 \\
\enddata
\end{deluxetable}

\begin{deluxetable}{l l c}
    	\centering
	\tablewidth{0pt}
	\tabletypesize{\scriptsize}
	\tablecaption{\label{tab:dr2_crossmatch} Crossmatch tables between DELVE DR2 and external catalogs.}
	\tablehead{
		\colhead{Column Name} & \colhead{Description} & \colhead{Columns}
	}
        \startdata
	DEC1 & Declination from DELVE DR2 (deg) & 1\\
	DEC2 & Declination from external catalog (deg) & 1\\
        DISTANCE & Angular separation between RA1,DEC1 and RA2,DEC2 (arcsec) & 1 \\
        ID1 & ID in DELVE DR2 (\var{quick\_object\_id}) & 1 \\
        ID2 & ID in external catalog (\var{source\_id}) & 1 \\
        RA1 & Right ascension from DELVE DR2 (deg) & 1 \\
        RA2 & Right ascension from external catalog (deg) & 1 \\
\enddata
\end{deluxetable}


\section{Depth}
\label{app:depth}

\begin{figure*}[t]
\centering
\includegraphics[width=0.8\textwidth]{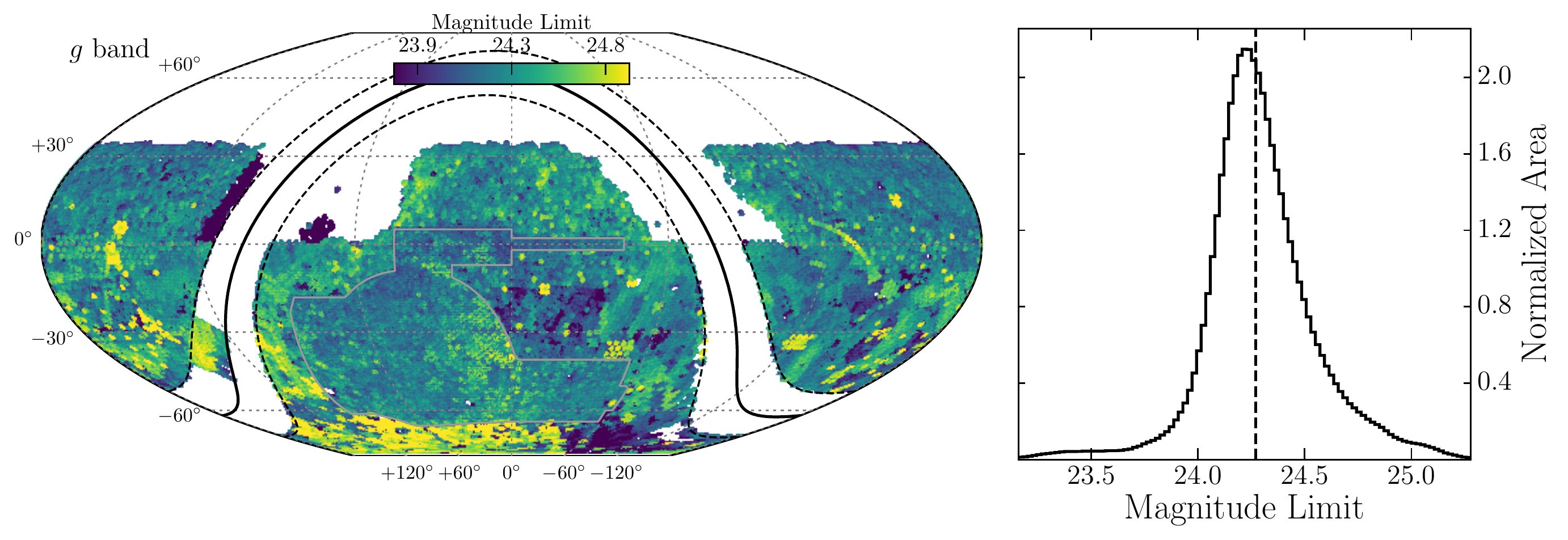}\\
\includegraphics[width=0.8\textwidth]{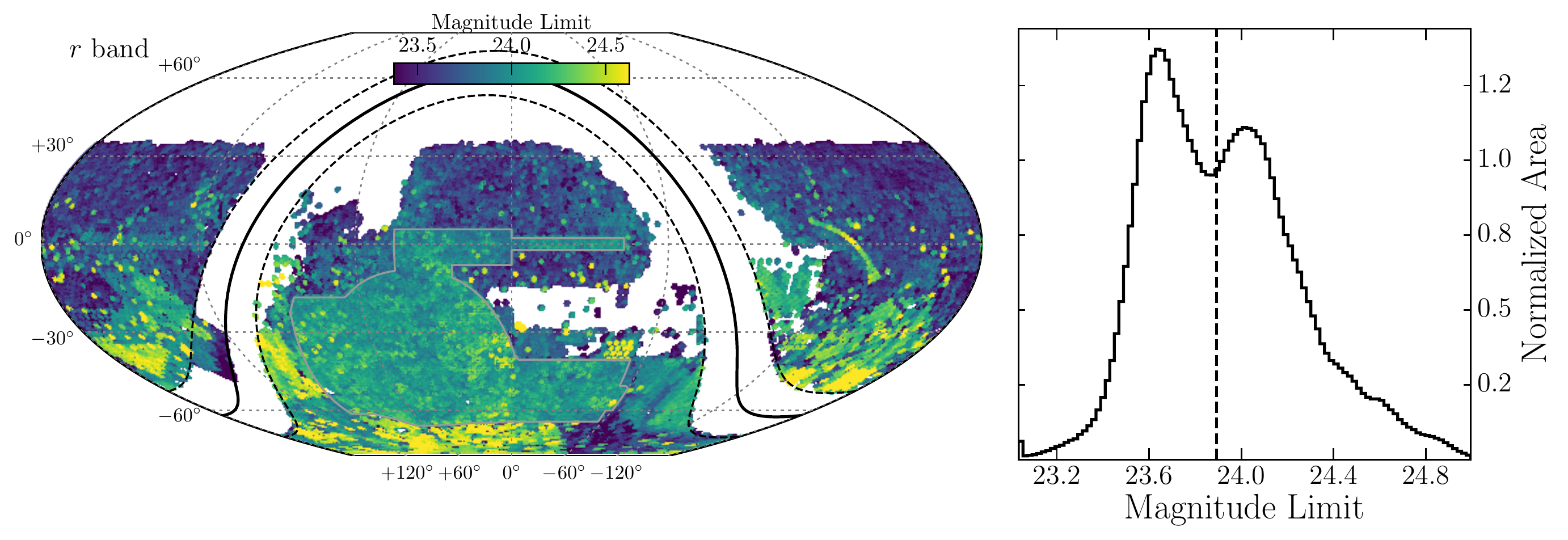}\\
\includegraphics[width=0.8\textwidth]{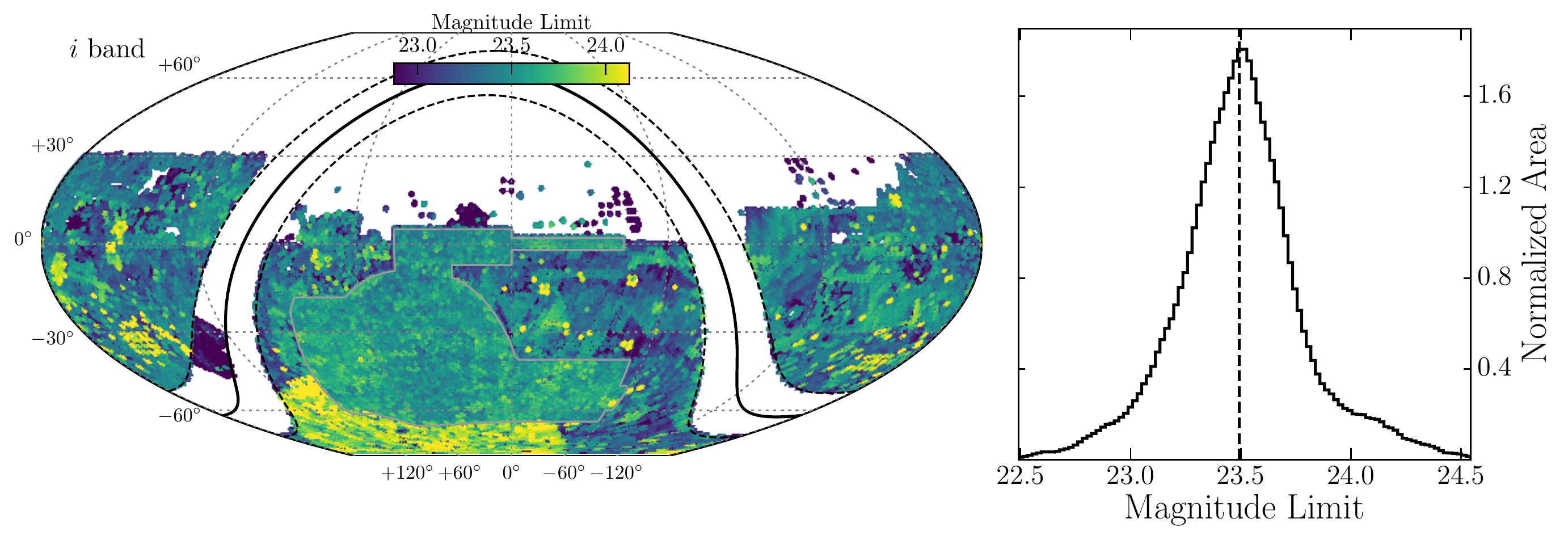}\\
\includegraphics[width=0.8\textwidth]{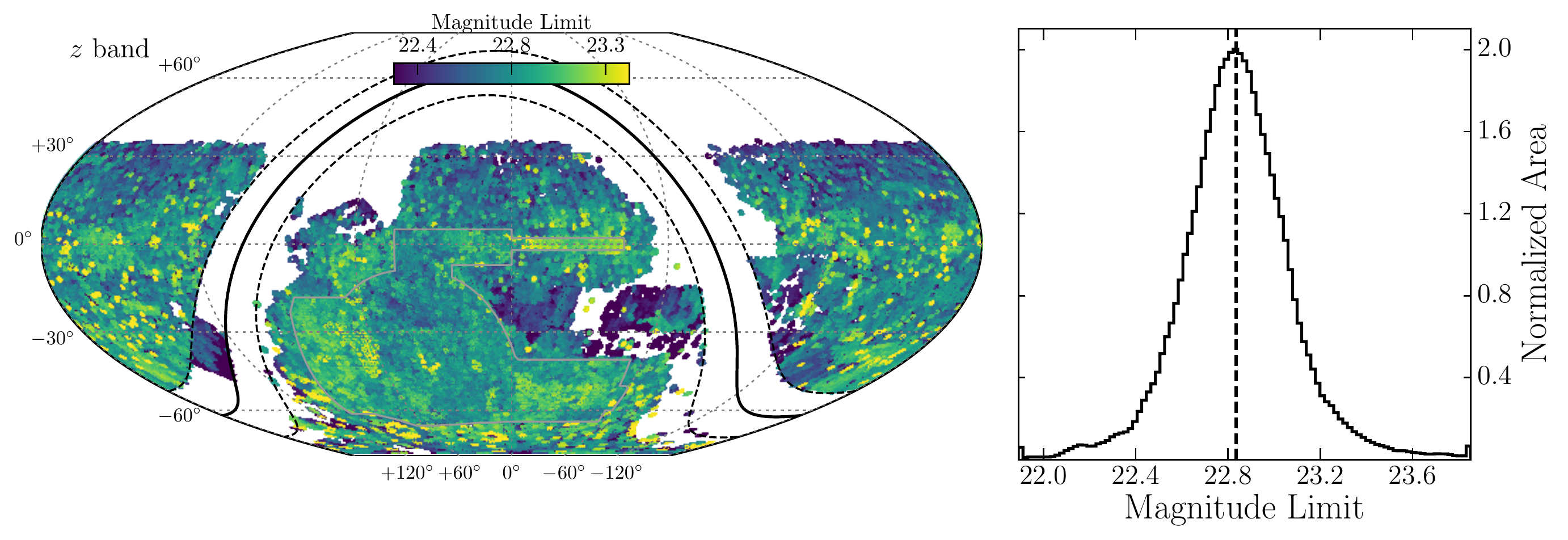}
\caption{Sky maps and histograms of the S/N=5 magnitude limit computed from the statistical uncertainty in \var{MAG\_PSF}. Dashed vertical lines indicate the median depth quoted in \tabref{summary}. Sky maps are plotted using an equal-area McBryde--Thomas flat polar quartic projection in celestial equatorial coordinates.
}
\label{fig:maglim_map}
\end{figure*}

This appendix includes sky maps showing variations in the S/N=5 depth of DELVE DR2 in the $g,r,i,z$ bands. The S/N=5 depth was derived from the magnitude at which the median magnitude uncertainty was $\delta m = 0.2171$ mag (\secref{depth}). These values were derived in  $\roughly 12 \amin^2$ \healpix pixels ($\nside = 1024$) and are shown in \figref{maglim_map}. 

\clearpage

\bibliographystyle{yahapj_twoauthor_arxiv_amp}
\bibliography{main}

\end{document}